\documentclass{article}

\usepackage{arxiv}

\usepackage[utf8]{inputenc} % allow utf-8 input
\usepackage[T1]{fontenc}    % use 8-bit T1 fonts
\usepackage[hidelinks]{hyperref}       % hyperlinks
\usepackage{url}            % simple URL typesetting
\usepackage{booktabs}       % professional-quality tables
\usepackage{amsfonts}       % blackboard math symbols
\usepackage{nicefrac}       % compact symbols for 1/2, etc.
\usepackage{microtype}      % microtypography
\usepackage{lipsum}
\usepackage{bm}
\usepackage{amsmath}
\usepackage{amssymb}
\usepackage{graphicx}
\usepackage{import}
\usepackage[rgb]{xcolor}
\usepackage[english]{babel}
\usepackage{natbib}
\title{Drop impact onto a substrate wetted by anotherliquid: Corona detachment from the wall film}

\author{ Bastian Stumpf\\
Institute for Fluid Mechanics and Aerodynamics, Technische Universität Darmstadt, Darmstadt, Germany\\
\texttt{stumpf@sla.tu-darmstadt.de} \\
\AND
Ilia V. Roisman \\
Institute for Fluid Mechanics and Aerodynamics, Technische Universität Darmstadt, Darmstadt, Germany\\
  \texttt{roisman@sla.tu-darmstadt.de} \\
  \AND
  Alexander L. Yarin\\
  Department of Mechanical and Industrial Engineering, University of Illinois at Chicago, Chicago, USA\\
  \texttt{ayarin@uic.edu} \\
  \AND
  Cameron Tropea\\
  Institute for Fluid Mechanics and Aerodynamics, Technische Universität Darmstadt, Darmstadt, Germany\\
  \texttt{ctropea@sla.tu-darmstadt.de} \\
}

\begin{document}
\maketitle
\begin{abstract}
Drop impact onto a thin liquid film of another liquid is observed and characterized  using a high-speed video system. A new mode of splash - a complete, simultaneous corona detachment - has been observed, which is the result of the lamella breakup near the wall film. The abrupt outward and upward displacement  of the lamella leads to an extreme stretching of the corona wall, resulting in rapid thinning and a rupture of the corona wall. This rupture triggers propagating Taylor-Culick rims, which rapidly spread, meet and thus undercut simultaneously the entire corona, resulting in its detachment. Special experiments with the spreading corona impingement onto a fixed needle, supplement the physical evidence of the above-mentioned mechanism. A self-consistent theory of the observed phenomena is proposed  and compared with experiments, exhibiting good agreement.
\end{abstract}

% keywords can be removed

\section{Introduction}
Splashing, resulting from drop impact onto a  film of another liquid is of high significance due to its importance in many industrial applications. For example, fuel mixture preparation and emissions in modern combustion engines are influenced by the interaction of fuel spray drops impacting onto lubricating oil films in the cylinder. Spray cooling during the process of hot forging \citep{yang2005physiothermodynamics} or functional printing \citep{layani2014printing} are further examples of technologies which involve drop/film interaction of different liquids. In these examples the drop/wall interaction is affected by the fact that the drop and the liquid film on the wall are different liquids and may exhibit different degrees of miscibility.

Among the quantities most often studied in the field of drop impact are the splashing threshold in terms of dimensionless impact parameters, diameter of the secondary droplets and their combined mass  compared to the mass of the impacting drop, as well as the diameter of the drop or of the corona. The functional dependencies of these quantities are described using the Weber number, We $= \rho D_0 U_0^2/\sigma$, the Reynolds number, Re $=\rho D_0 U_0/\mu$, (or their combination) and the dimensionless initial wall film thickness, $\tilde{\delta}=H_\text{f0}/D_0$, where $U_0$ is the drop impact velocity, $D_0$ is the drop diameter and $H_\text{f0}$ is the initial wall film thickness.  Comprehensive reviews  of drop impact phenomena and their modeling can be found in the literature \citep{yarin2006drop,josserand2016drop,marengo2011drop,CollisionPhenomena2017}. 

The understanding of the dynamics of drop impact onto a liquid film is based on the seminal work of \cite{yarin1995impact}, where an asymptotic solution for an inviscid wall flow is found and the splash phenomena are described as a propagation of a kinematic discontinuity in a spreading liquid film on the wall. This study proposes a reliable and widely accepted form for the description of the splashing threshold, which depends, among other parameters, on the drop impact frequency. Following this theory, the outward  lamella velocity $\tilde{u}$, the  lamella thickness $\tilde{h}$ and the  radius of the corona $\tilde{R}_c$ can be expressed in  dimensionless form as
\begin{equation}\label{eq:yarin}
   \tilde{u}=\frac{\tilde{r}}{\tilde{t}+\tilde{\tau}},\quad \tilde{h}=\frac{\eta}{(\tilde{t}+\tilde{\tau})^2}, \quad \tilde{R}_c=\beta \sqrt{\tilde{t}+\tilde{\tau}}.
\end{equation}
where $\tilde{\tau}$ and $\beta$ are constants. Here the drop initial diameter is used as a length scale, the impact velocity as a velocity scale, and their ratio as a time scale.

Propagation of the corona in a liquid film has been studied in \cite{roisman2008propagation}. Experiments show that surface tension influences the evolution of the corona radius, leading to its deviation from the predicted square-root dependence of $\tilde{R}_c$ on time, Eq.~(\ref{eq:yarin}). Moreover, surface tension and gravity lead to a receding of the impact crater in the film fluid  after its diameter reaches the maximum value $D_\mathrm{max}$. 
%As shown in Figure~\ref{img:DmaxTmax}, t
The dimensionless maximum crater diameter and the corresponding dimensionless spreading time $\tilde{t}_\mathrm{max}$  are only slightly dependent on the initial film thickness and on the liquid viscosity, and are determined mainly by the Weber number. 

%\begin{figure}
%	\centering
%	\includegraphics[width=0.49\textwidth]{HinsbergDmax.eps} \includegraphics[width=0.49\textwidth]{HinsbergTmax.eps}
%	\caption{Dimensionless maximum crater diameter $D_\mathrm{max}$ and the corresponding time $T_\mathrm{max}$ as a function of the Weber number for different  relative film thicknesses $0.5 \leq \delta \leq 2$. The data are from \cite{van2010investigation}. The time of the drop initial penetration into the wall film is excluded from the total time to estimate the time of corona spreading $T_\mathrm{max}$ as explained in \cite{roisman2008propagation}.  The Reynolds numbers are in the range $4700 < Re < 20200$. Water and isopropanol are used in the experiments of \cite{van2010investigation} to vary the surface tension. }
%	\label{img:DmaxTmax}
%\end{figure} 

Many subsequent studies have confirmed these relations experimentally \citep{cossali1997impact,bakshi2007investigations}. These relations are valid for drop spreading on a dry solid substrate or for drops impacting onto a thin liquid film. However, it should be noted that these expressions are valid only for cases when the film thickness is much thicker than the viscous boundary layer, which would be  formed by the spreading lamella at the wall surface. The viscous boundary layer leads to a damping of the spreading lamella and to formation of a residual film \citep{yarin1995impact,roisman2009inertia}.

Among the  well-studied outcomes of drop impact onto a wetted substrate are drop deposition, drop bouncing and corona splash. Which of these phenomena occur is determined by inertial, viscous and capillary forces. The phenomenon of splash is one of the most important phenomena because of being central  in many industrial applications, especially due to the liquid mass which does not remain on the surface, but is rather ejected in the form of a corona and, finally, secondary droplets.

Two main types of splashing have been observed upon drop impact \citep{worthington1897impact,harlow1967splash,levin1971splashing,macklin1976splashing,wang2000splashing,rioboo2001outcomes}: prompt splash and corona splash. Under certain conditions also corona detachment has been observed upon spray impact in microgravity,  \cite{roisman2007breakup}, leading to formation of larger secondary droplets formed from the corona rim. 

Empirical correlations were formulated for the splashing threshold
\begin{eqnarray}\label{eqCoss}
K&=&2100-2700 \exp(-58 \tilde{\delta}),\quad  0.02 <\tilde{\delta} < 0.1,\label{eqCoss1}\\
K&=&2100+5880 \tilde{\delta}^{1.44},\quad  0.1 <\tilde{\delta} < 1.\label{eqCoss2}
\end{eqnarray}
based on the experiments of \cite{rioboo2003experimental} and \cite{cossali1997impact}, where the splashing threshold parameter $K$ is defined as
\begin{equation} \label{eq:K}
   K\equiv\mathrm{We}^{4/5} \mathrm{Re}^{2/5}.
\end{equation}
One important phenomenon is the rupture of the liquid film on a wall as a result of the fast spreading produced by drop or spray impact \citep{kadoura2013rupture}. The flow instability leading to the film rupture can be enhanced by the presence of the second liquid of different surface tension. This phenomenon could potentially  increase the wall deposition ratio significantly, for example,  under  conditions typical of those found in internal combustion engines.

Several studies focused on the investigation of complex liquid drop impact, for example suspensions or emulsions \citep{prunet1998impacting,bolleddula2010impact,hao2016dynamic,derby2010inkjet,derby2003inkjet}, or encapsulated drops \citep{chiu2005experiment} onto dry substrates. One recent study by \cite{lhuissier2013drop} is devoted to the impact of a liquid drop onto a deep pool of another immiscible liquid. It was shown that at some threshold velocity the impact leads to drop disintegration  into several fragments and thus, to the liquid emulsification. Similar phenomena could occur  after drop impact onto a liquid film, if the impact velocity is high enough. 

In \cite{Kittelsplash2018} the splashing threshold of a liquid drop impacting onto a solid substrate wetted by another liquid was studied. Three main regions are identified. For the case when the kinematic viscosity of the wall film ($\nu_f$) is much higher that the kinematic viscosity of the drop ($\nu_d$), the properties of the drop govern the process of splashing. For  the case of drop kinematic viscosity much higher than that of the film, the splash is governed by the properties of the film. For the case when the kinematic viscosities of the film and  the drop are comparable, the splashing threshold depends on the viscosity ratio, defined as $\tilde{\kappa}_\nu=\nu_f/\nu_d$. 

In the present experimental study the impact of a liquid drop onto a solid substrate wetted by another liquid is investigated. A range of impact parameters and combinations of liquid properties have been determined for which the corona produced by the impact completely and simultaneously detaches from the film. \textcolor{black}{ The corona detachment is a mode of  splashing which leads to the generation of large secondary drops} %\textcolor{red}{The corona detachment is a mode of splashing that can superimpose on other drop impact outcomes, such as corona splash or corona formation without rim instability,  leading to the formation of additional secondary droplets.} 
This phenomenon has been  observed before \citep{roisman2007breakup,LamannaGeppert2016}, but the mechanisms of corona detachment and the conditions leading to this phenomenon have not been conclusively analyzed or understood. The present work contains such an analysis and provides a self-consistent theory of this phenomenon.

\section{Experimental method}
\begin{figure}
\small
\centering
	\includegraphics[width=0.6\textwidth]{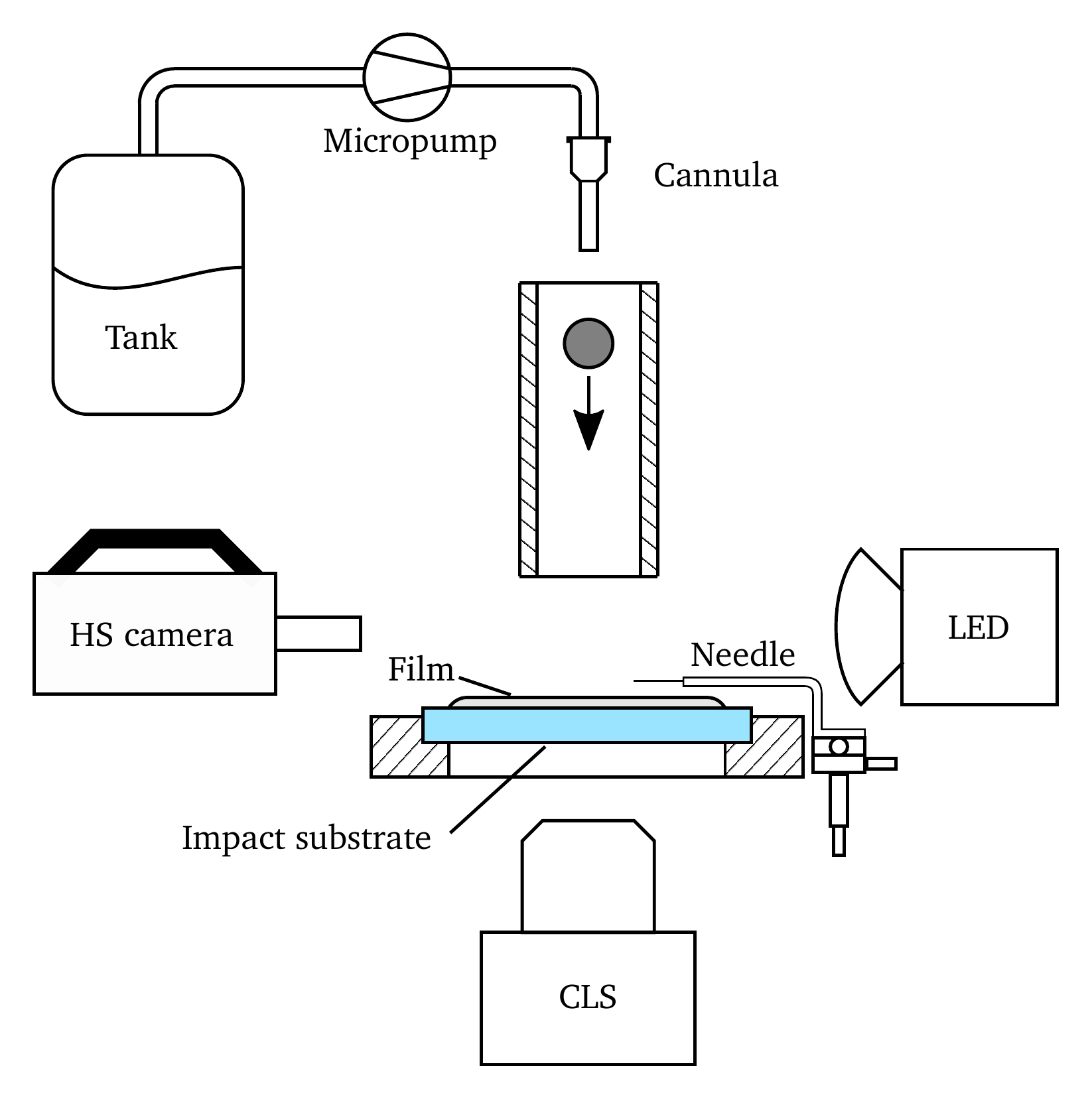}
\caption{Schematic representation of the experimental setup}
\label{fig:ExperimentalSetup}
\end{figure}

%\subsection{Experimental setup}
The experimental setup, shown in Figure~\ref{fig:ExperimentalSetup}, consists of three parts, the drop generator, the impact substrate and the observation system. The drop generator is based on a drop-on-demand design. A micropump transports  fluid from a reservoir tank to the cannula. The fluid forms a drop at the  tip of the cannula. The drop drips off the cannula tip by gravity, once a critical mass is reached. 
The initial drop diameter ($D_0$) is varied in this study from 1.45 mm to 2 mm. 
The impact velocity of the drop ($U_0$) is varied by varying the height of the cannula tip above the impact substrate.

The impact substrate is a round, horizontally aligned sapphire glass plate with a diameter of 50~mm and a height of 0.5~mm.The plate is made of sapphire glass, which is optically polished to minimize the effect of surface roughness. The film thickness is measured using a chromatic-confocal line (CL) sensor (Precitec CHRocodile CLS). This system allows the film thickness to be measured at 192 points on a line of 4.5~mm length. In order to lay a film of a defined thickness, the film fluid is applied at the center of the sapphire glass plate where it spreads. The film thickness is then varied from 28~$\mu\text{m}$ to 120~$\mu\text{m}$ by utilizing a spin coating process in which the film thickness is constantly monitored by the CL-sensor. The fluids used are silicone oils with varying kinematic viscosity,
as shown in Table~\ref{Tab:1}.

\begin{table}
 \begin{center}
\begin{tabular}{p{1cm}ccc}
  Fluid & Kinematic viscosity & Surface tension & Density\\
  & \(\nu\) [mm$^2$/s] & \(\sigma\) [mN/m] & \(\rho\) [kg/m$^3$] \\ 
  \hline

     S5 & 5  &17.72 & 920\\
     S10 & 10 &18.29 & 930\\
     S20 & 20 &18.2 & 945\\
%     S25 & 25 &18.19\\
%     S50 & 50 & 18.69\\
%     S65 & 65 &18.69\\
%     S750 & 750 &18.78\\
\end{tabular}
\caption{Fluid properties. Sxx - for  silicone oils of different viscosity.}\label{Tab:1}
 \end{center}
\end{table}

The observation system consists of a high-speed video camera and an illumination source. The frame rate of the high-speed video camera (Photron Fastcam SA-X2) is set to 50\,000~fps with a resulting resolution of 768$\times$328 pixel and a shutter speed of 18.4~$\mu$s. This high framerate allows to precisely determine the instant of detachment. A light-emitting diode (Veritas Constellation 120E, 12\,000 Lumen) is used as an illumination source. A diffusing screen is placed in front of the illumination source to provide a uniform back lighting

An additional set of experiments is conducted in which an artificial rupture is induced in the liquid corona sheet by a needle. These measurements are aimed at observing the velocity of the rupture rim and to draw conclusions about the thickness of the corona sheet, as  elaborated in section~\ref{sec:experimental_characterisation}. The needle can be precisely positioned by a three-axes system so that it punctures the corona while it grows beyond the position of the needle. 

\begin{figure}
\small
\centering
	\includegraphics[width=\textwidth]{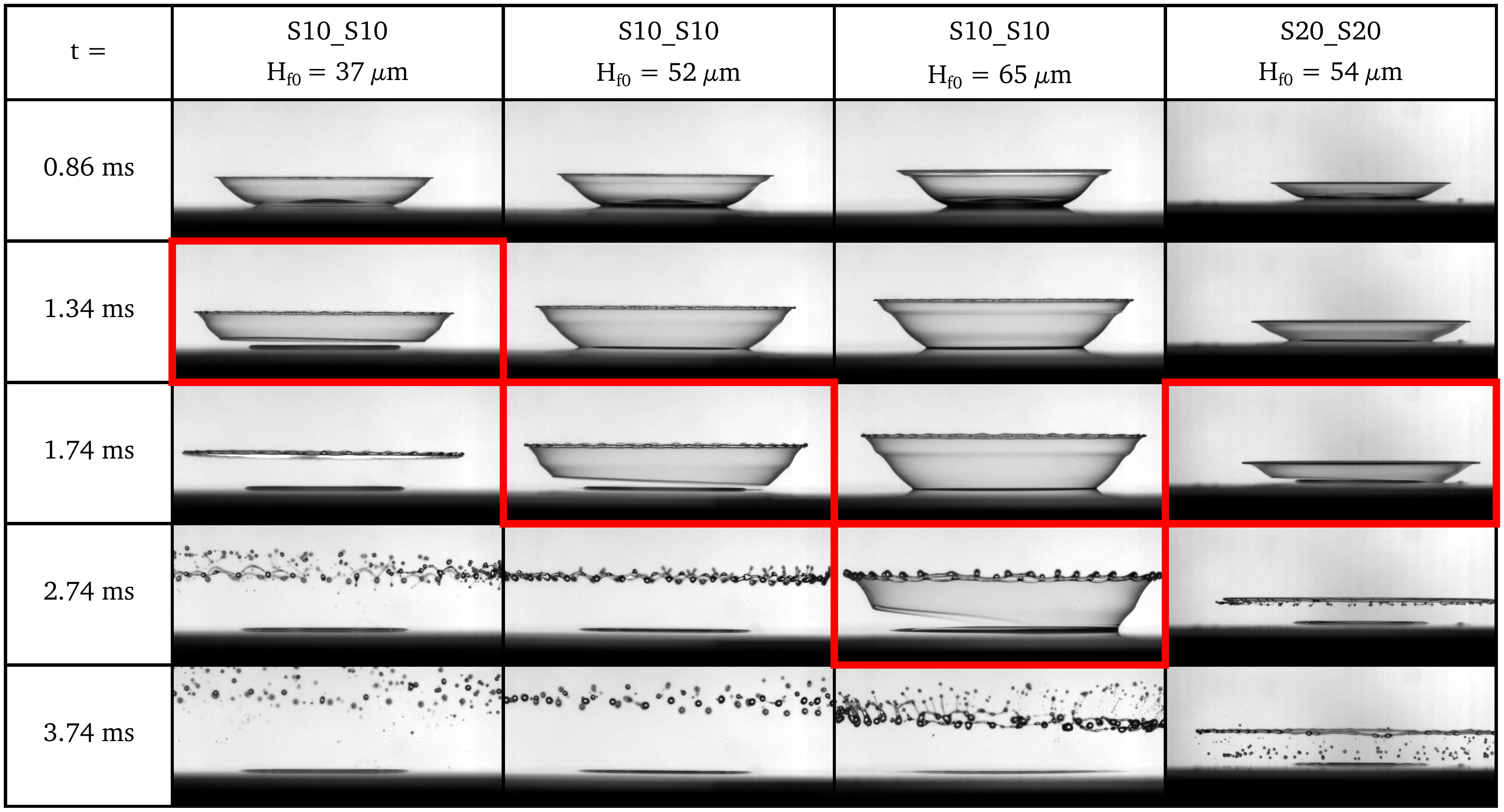}
\caption{Evolution of corona formation, detachment and atomization for cases of varying $H_{\mathrm{f0}}$ and viscosity. The instant shortly after detachment  is marked in each case with a red box. Impact parameters are: $D_0=2~\mathrm{mm}$, $U_0=3.2~\mathrm{m/s}$ and $\tilde{\kappa}_\nu=1$.}
%\textcolor{red}{The subscript f0 in the top labels should not be italics}}
\label{fig:corona_images}
\end{figure}

\section{Experimental characterization of the rupture process}
\label{sec:experimental_characterisation}
Typical outcomes of drop impact onto a liquid film are shown in Figure~\ref{fig:corona_images} for four  cases in which the drop and wall film are the same fluid but with varying film thickness and/or fluid viscosity. In all cases a corona evolves after impact and in all cases the corona detaches from the wall film, eventually retracting to the upper Taylor rim and disintegrating into ligaments and/or drops. The frame immediately after detachment is marked in this figure with a red box. In the present study attention is focused on the reason for this observed corona detachment. 

\begin{figure}
	\centering
		\includegraphics[width=0.7\textwidth]{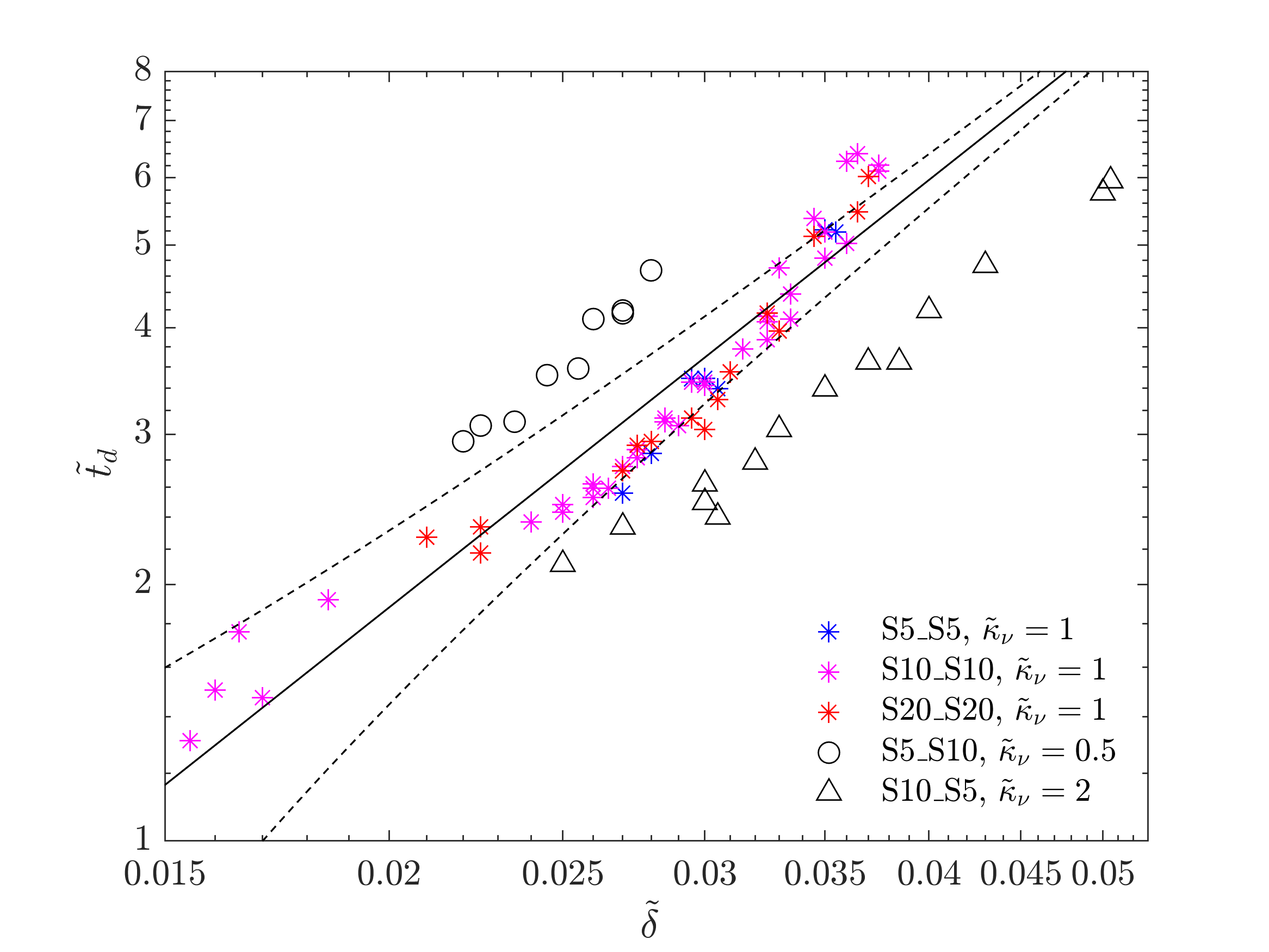}
	\caption{Dimensionless instant of corona detachment $\tilde{t}_d$ at different initial dimensionless wall film thickness $\tilde{\delta}$ and different fluid combinations. The fluid combinations are specified by  Sxx$\_$Syy (Sxx -- film fluid  and  Syy -- drop fluid). Impact parameters are $U_0 = 3.2$~m/s, $D_0 = 2$~mm. The time $\tilde{t}_d$ is rendered dimensionless using $D_0/U_0$. \textcolor{black}{The solid line represents the slope of Equation~(\ref{eq:detachmentmenttimefilmthicknessexp1}) with $k$=80.84 (obtained from the least mean-square method applied to the data from like fluids). The dashed lines indicate the range of one standard deviation  around the solid line.  }
     	 }
	\label{img:S2}
\end{figure}

While Figure~\ref{fig:corona_images} presents only exemplary observations, experiments have been carried out over a large range of wall film thickness and wall/film fluid combinations. At the high frame rate of the camera the first instant of detachment can be determined very accurately and furthermore, the critical wall thickness can be determined, beyond which corona detachment can no longer be observed ($\delta_{\textrm{crit}}$). The results of these experiments are summarized in Figure~\ref{img:S2} and Table~\ref{tab:deltacrit}.

In Figure~\ref{img:S2}  the dimensionless time of detachment $\tilde{t}_d$  (made dimensionless with $D_0/U_0$) is shown  as a function of the initial dimensionless wall film thickness $\tilde{\delta}$   for different combinations of the drop and film liquid viscosities.
 It is interesting to note that in the cases with the same liquid in the drop and wall film, the viscosity does not  have a significant influence on $\tilde{t}_d$, since the results for S5\_S5,  S10\_S10 as well as S20\_S20 overlap closely.
  From this figure it is also apparent that for the different fluid combinations, corona detachment can only be observed for some maximum dimensionless wall film thickness, beyond which no data points are shown. This limiting dimensionless wall film thickness, $\tilde{\delta}_{\textrm{crit}}$, is summarized in Table~\ref{tab:deltacrit} for the investigated fluid combinations. While the critical dimensional wall film thickness for corona detachment   remains constant for a single component drop impact with the fluids S5 and S10, the critical value of $H_\text{f0}$ is higher for S20. This can be explained by the fact that in the S5\_S5 and  S10\_S10 cases the corona detachment is superimposed on a crown splash and at the instant $\tilde{t}_d=5.8$ when the corona in the S20\_S20 case detaches, the corona in the other two cases has already collapsed due to the rim instability and can, therefore, no longer detach. 
  
  \begin{table}
\centering
%\vspace{6pt}
\begin{tabular}{c|ccccc}
%\toprule
Fluids  & S5\_S5 & S10\_S10  & S20\_S20 & S5\_S10 & S10\_S5\\
  $\tilde{\kappa}_\nu $  &  1   &   1   &       1      &   0.5   &   2.0  \\
\hline 
$\tilde{\delta}_{\textrm{crit}}$ & 0.0375 & 0.0375 & 0.04& 0.03&$>$0.06\\
$H_\text{f0}$ [$\mu$m] & 75 & 75 & 80 & 60& $>$120\\  
%\bottomrule 
\end{tabular}
\caption{Maximum dimensionless wall film thickness for which detachment can be observed $\delta_{\textrm{crit}}$ for different fluid combinations. $ U_0=3.2 $m/s, $ D_0=2 $mm. Sxx\_Syy specifies fluid-drop fluids; $\tilde{\kappa}_\nu = \nu_f/\nu_d $}
\label{tab:deltacrit}
\end{table}

  Finally, the data in Figure~\ref{img:S2} reveals a rather strong influence of the viscosity ratio  $\tilde{\kappa}_\nu$, exhibiting significantly longer dimensionless detachment times for higher values of $\tilde{\kappa}_\nu$.
  
\begin{figure}
	\centering
	\includegraphics[width=0.8\textwidth]{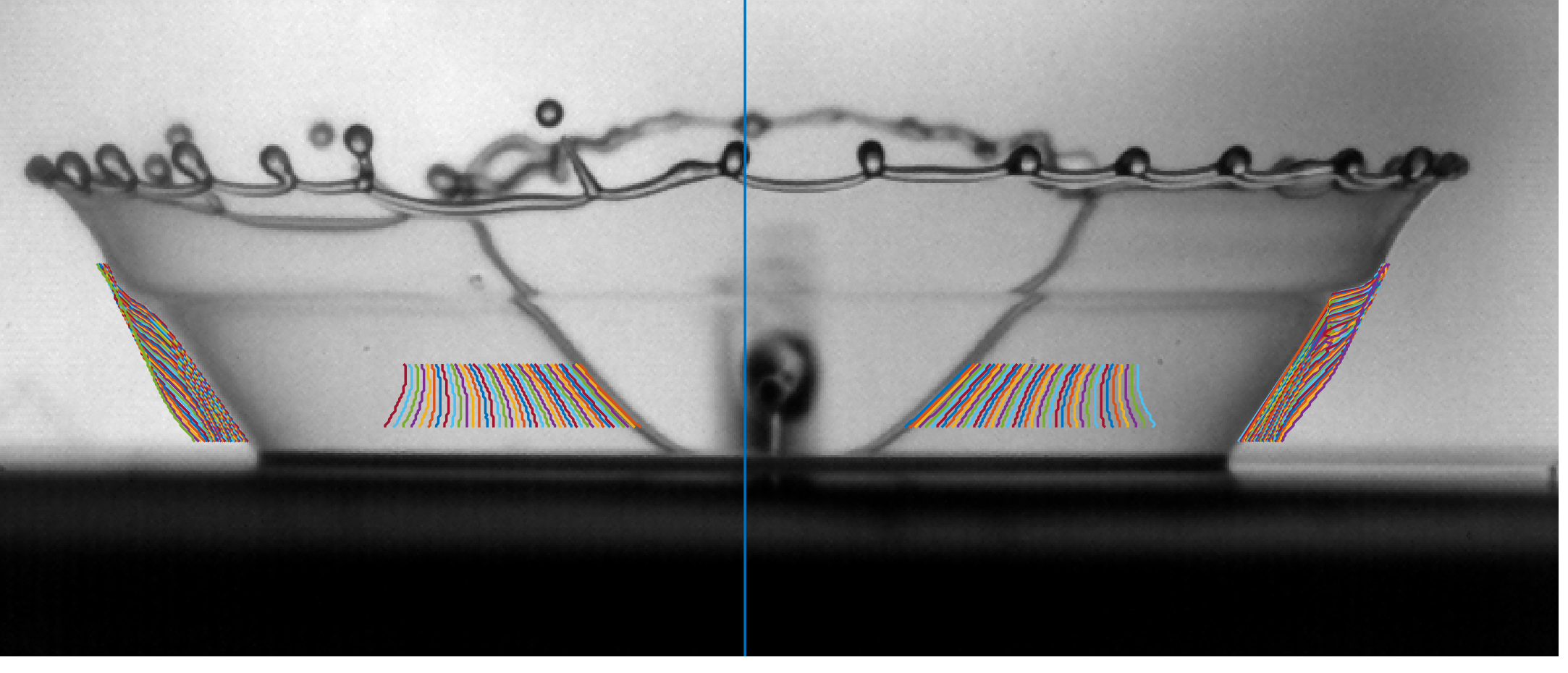}
	\caption{ Drop impact at $t=2.6\text{ms}$. Blue line denotes corona centerline/axis. The rim contour at different subsequent time instants is shown with different colours on the left and right side of the centerline. Impact parameters: $D_0=2\mathrm{mm}$, $U_0=3.2\text{m/s}$, $H_{\text{f0}}=80~\mu \text{m}$}
	\label{img:CPRup}
\end{figure}

To determine the thickness of the corona liquid sheet at the instant of corona detachment, the artificial rupture with the needle was used. This technique is visualized in Figure~\ref{img:CPRup}. In this figure the needle is held horizontal to the surface and pierces the liquid corona sheet on the far side as it expands. The needle can be positioned such that the puncturing takes place just prior to an expected corona detachment.  This initiates a hole, which then propagates throughout the sheet. By tracking the Taylor rim contour in time, shown in the figure with coloured lines, the velocity of the rim can be computed as a function of height above the surface and time. Each contour line comes from a frame subsequent to the frame shown in the image. The axisymmetric center of the corona is indicated with a blue line. This velocity can then be used to compute the film thickness, thus providing film thickness information for the corona sheet just prior to corona detachment.

The  thickness in the rupturing corona film can now be estimated with the help of the Taylor-Culick mechanism \citep{Taylor1959,Culick1960} describing the relation between surface tension $\sigma$, density $\rho$, rim velocity $u_{\textrm{TC}}$ and corona wall thickness $h$ in a tearing corona  as     
 
\begin{equation}
    \label{eq:Taylor}
    h=\frac{2\sigma}{u_{\textrm{TC}}^2\rho}
\end{equation}

However, the rim does not move on a planar liquid sheet, but along the conical contour of the corona. Thus, the horizontal displacement of the rim obtained from the high-speed video images must first be projected onto a  path laying on the corona contour. Assuming that the corona is rotational symmetric, the angle through which the rim propagates between two consecutive times $t_1$ and $t_2$ is given by 

\begin{align}
\label{eq:RimRadiant}
    \Delta \alpha =  \bigl| \arcsin \biggl( \frac{x(t_1,y)}{r(t_1,y)} \biggl) -\arcsin \biggl({\frac{x(t_2,y)}{r(t_2,y)}}\biggr) \bigr|, 
\end{align}
\noindent where $x$ is the horizontal distance measured between consecutive contour lines projected onto the image plane (as shown in Figure~\ref{img:CPRup}), and  $r$ is the radial distance from the corona center axis to the corona sheet. Both quantities are a function of time and height above the corona base, $y$.  These two quantities can be obtained directly from the high-speed images, as is is indicated in Figure~\ref{img:CPRup}, where the outline of the rim and the corona border in subsequent time steps is plotted as coloured lines. 

To compute the velocity of the rim in the horizontal direction, $u_m$, the average radius of the corona is used, i.e., 

\begin{align}
\label{eq:Um}
   \Delta s_m &= \bar{R} \Delta \alpha; \quad \bar{R}=\frac{R_1+R_2}{2}; \quad    u_m =\frac{\Delta s_m}{\Delta t}  
\end{align} 
\noindent where $R_1$ and $R_2$ are the corona radii at the times $t_1$ and $t_2$ and $\Delta t = t_2 - t_1$.

\begin{figure}
\begin{minipage}{.45\textwidth}
	\centering
		\includegraphics[width=0.9\linewidth]{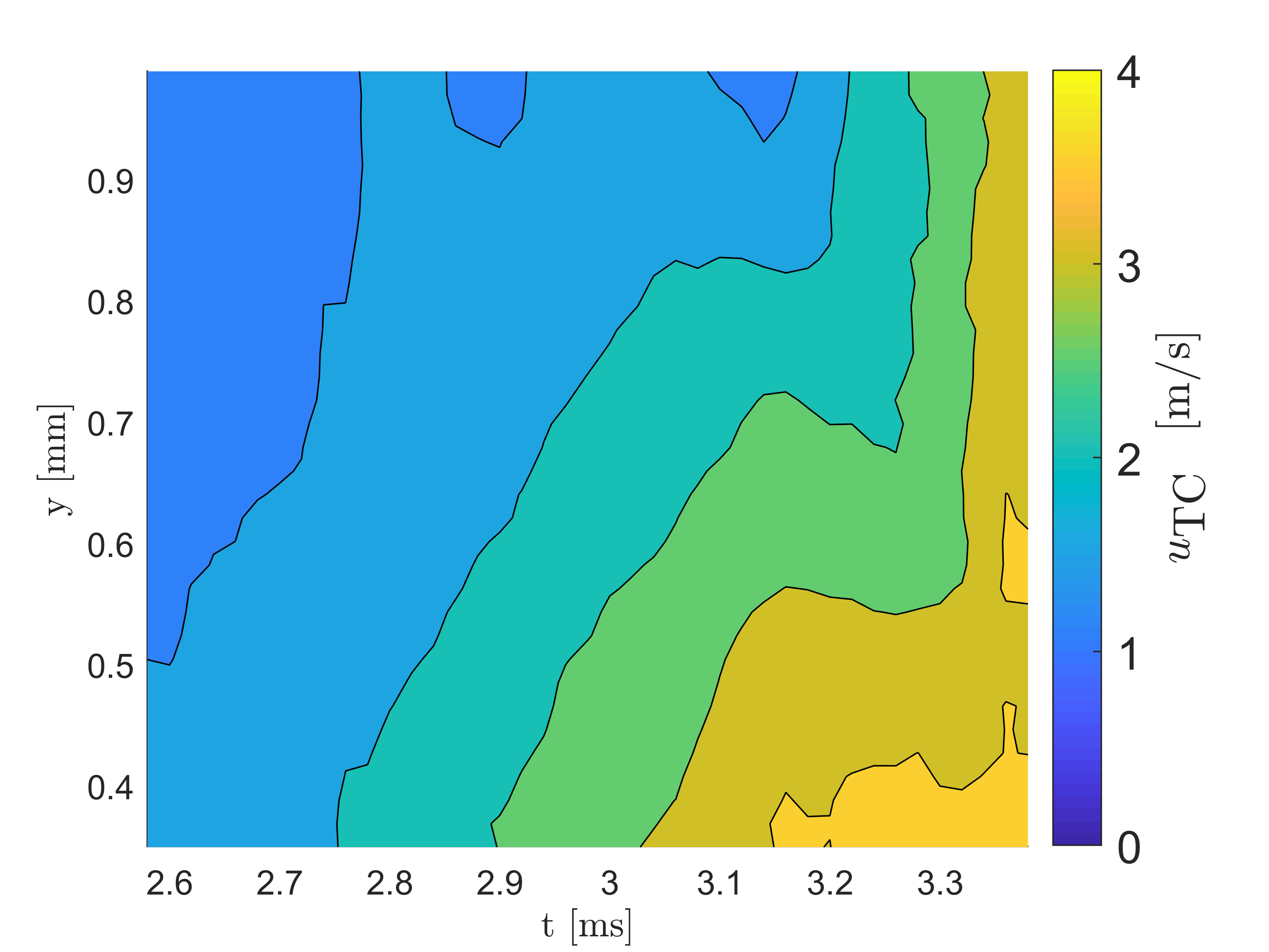}
	%\captionof{figure}{ test } 
	\label{img:UFitL}
\end{minipage}
\hspace{.08\linewidth}
\begin{minipage}{.45\textwidth}
		\includegraphics[width=0.9\linewidth]{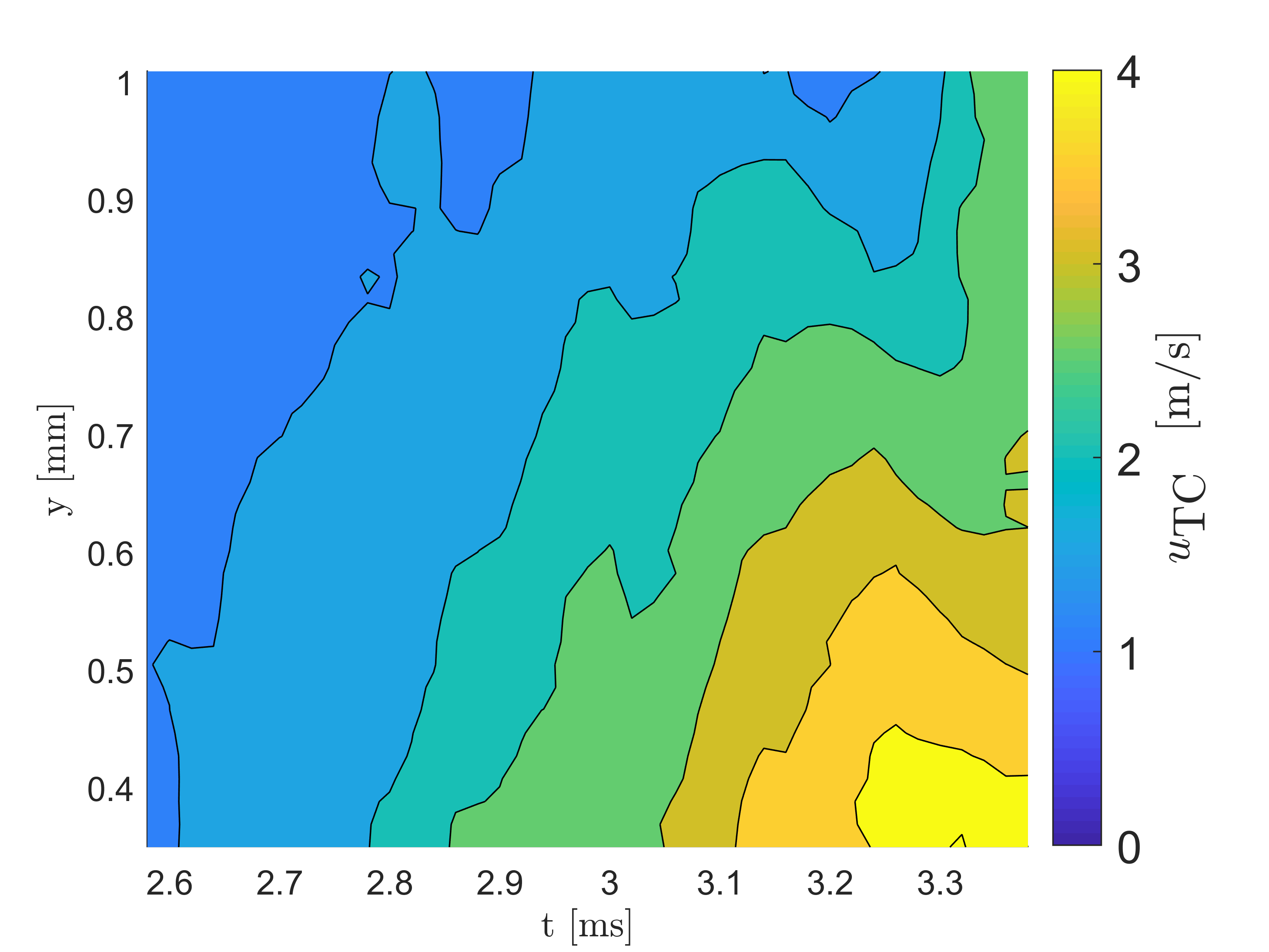}
	%\captionof{figure}{ }
	\label{img:UfitR}
\end{minipage}
\caption{Measured rim velocity  to the left and right side of the corona centerline. Impact parameters: $D_0=2$mm, $U_0=3.2\text{m/s}$, $H_{\text{f0}}=80~\mu \text{m}$. The velocities on both sides are shown as positive values for better comparison.  }
\label{img:Ufitfilt}
\end{figure}

\begin{figure}
\begin{minipage}{.45\textwidth}
	\centering
		\includegraphics[width=0.9\linewidth]{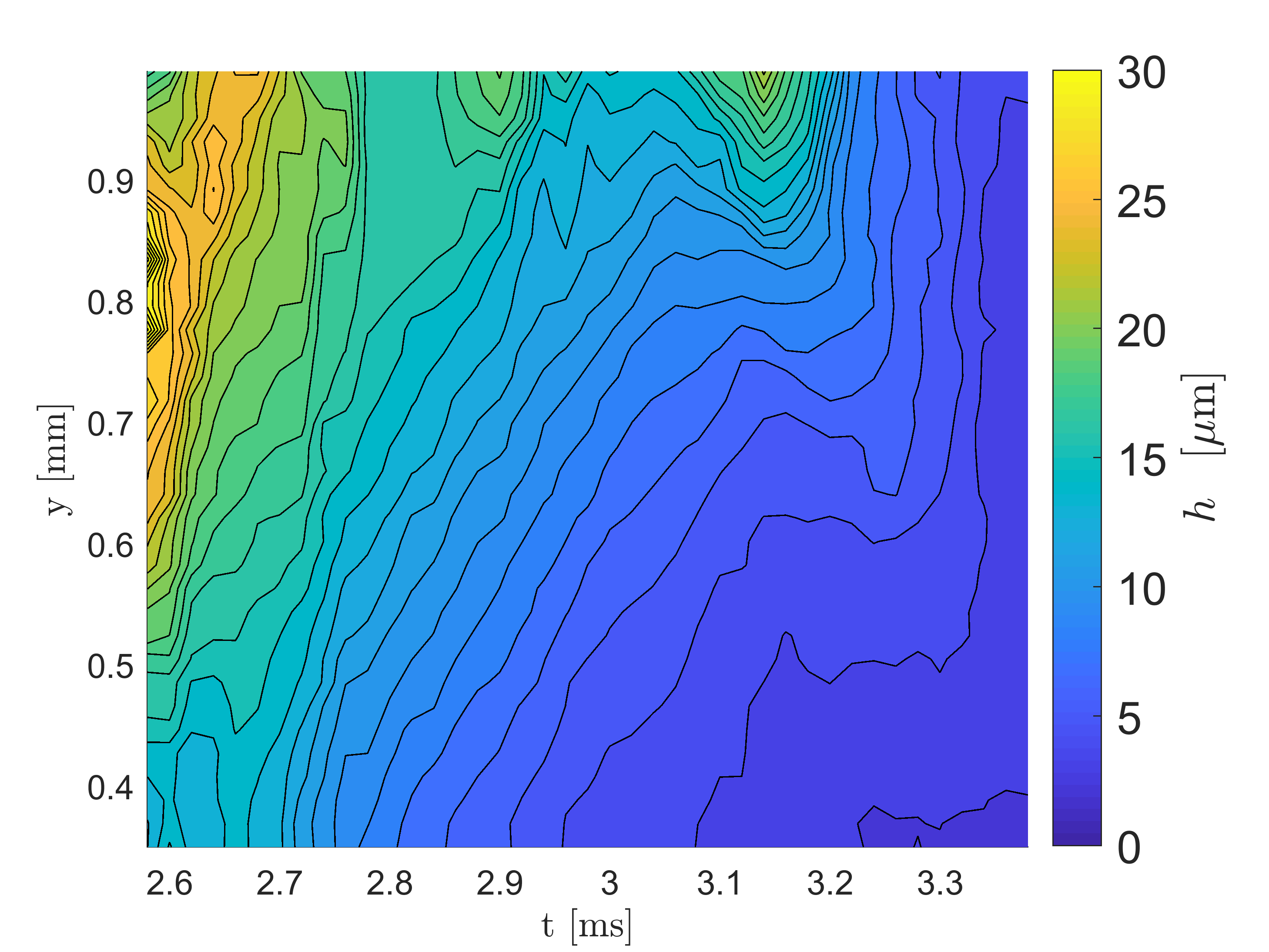}
	%\captionof{figure}{ test } 
	\label{img:hcFitL}
\end{minipage}
\hspace{.08\linewidth}
\begin{minipage}{.45\textwidth}
		\includegraphics[width=0.9\linewidth]{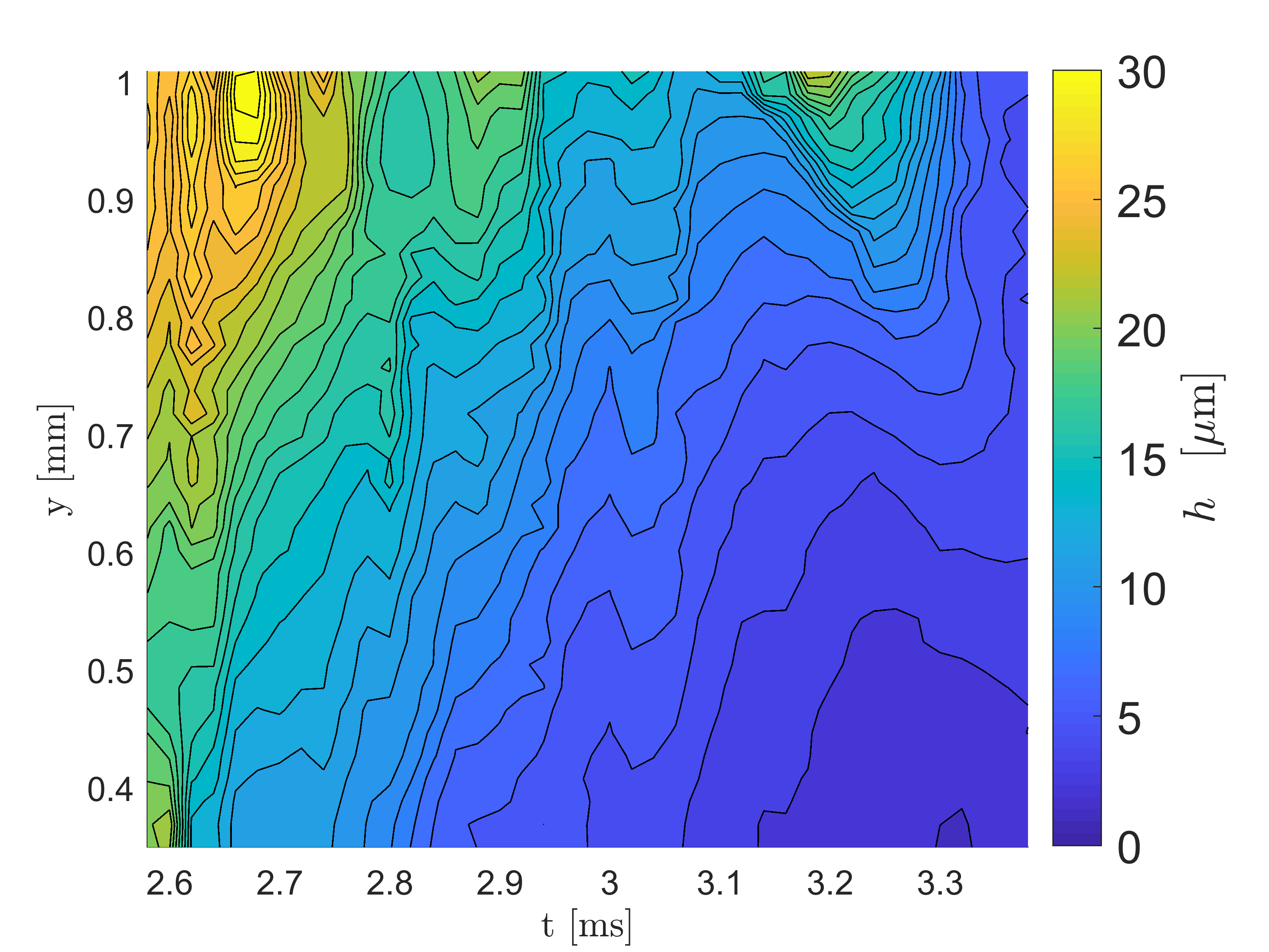}
	%\captionof{figure}{ }
	\label{img:hcfitR}
\end{minipage}
\caption{Film thickness on the left and right side of the corona centerline calculated from Eq.~(\ref{eq:Taylor}) using the velocities shown in Figure~\ref{img:Ufitfilt}. Impact parameters: $D_0=2$mm, $U_0=3.2\text{m/s}$, $H_{\text{f0}}=80~\mu \text{m}$. }
\label{img:hcfitfilt}
\end{figure}

The velocity $u_{\textrm{TC}}$ appearing in Eq.~(\ref{eq:Taylor}) is the velocity normal to the rim, which would require following material points in the sequence of images available from the high-speed camera. Since this is not possible, the velocity $u_{\textrm{TC}}$ was estimated by using the measured horizontal velocity of the rim, $u_m$, and the local inclination angle of the rim to the vertical, $\theta$, in the form $u_{\textrm{TC}}=u_m\cos{\theta}$. These computed velocities and film thicknesses are shown as a function of height above the corona base and of time in Figs.~\ref{img:Ufitfilt} and \ref{img:hcfitfilt}. The symmetry exhibited by these contour plots between the left and right side propagation of the rim is very high.

%\clearpage

\begin{figure}
	\centering
	\includegraphics[width=1\textwidth]{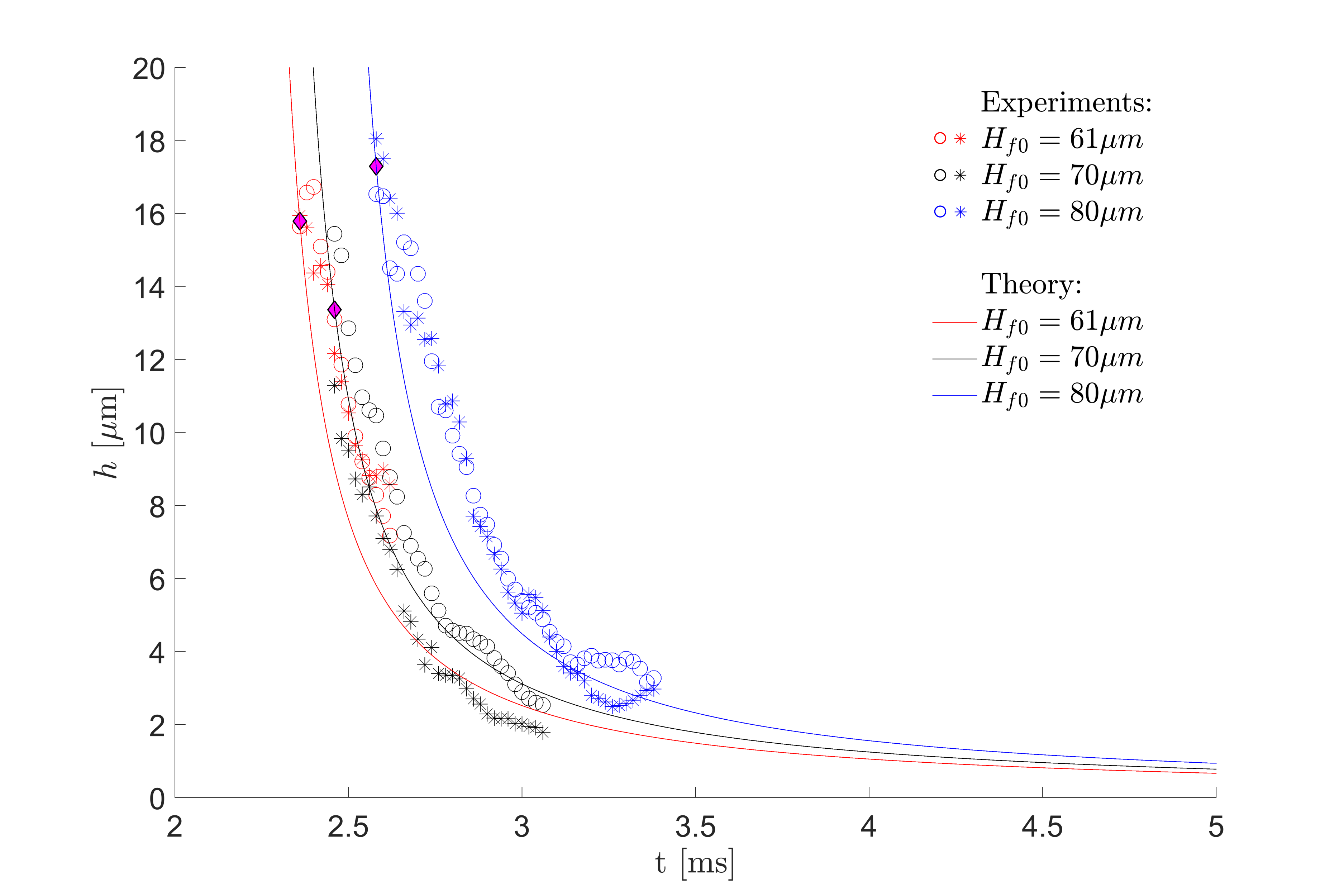}
	\caption{Comparison of temporal development of corona sheet thickness from experiments and theory. Experimental data shows the film thickness 0.46 mm above corona base.\textcolor{black}{The fluid for film and drop is S10 for all experiments.} $H_\text{f0}$ is varied from 60 to 80~$\mu$m and is colour coded (red: $H_\text{f0}=61\mu \text{m}$; black $H_\text{f0}=70\mu \text{m}$; blue $H_\text{f0}=80 \mu \text{m}$). The symbols (o and $\ast$) denote the left and right rim respectively.  The solid lines show the evolution of corona sheet thickness predicted by Eq.~(\ref{eq:filmthicknesscrownwalldim}), whereby $B$ is chosen to be 4. The temporal offset $t_0$ is \{2.31, 2.37, 2.51\}~ms for \{61, 70, 80\}~$\mu \text{m}$ \textcolor{black}{. The magenta coloured diamonds mark the film thicknesses and instances which were used to calculate the offset times  $t_0$. Since the film heights determined from the right- and left-hand sides of the  rim coincide well,  the mean value of $h$ from both sides at the first measured instance of each experiment has been chosen for reference. } 
	    }
	\label{img:hc_exp_theory}
\end{figure}

The corona sheet thickness 0.46~mm above the corona base is then plotted in Figure~\ref{img:hc_exp_theory} for three cases of substrate film thickness, $H_\text{f0}=61 \mu \text{m}$, 70$\mu \text{m}$ and 80$\mu \text{m}$. The observed trends are clear: the corona sheet reduces rapidly in thickness from about 18 $\mu$m to 3-4$\mu$m, or until detachment occurs.  For reference, in the case of $H_\text{f0}=61 \mu \text{m}$ the corona detachment occurs at 2.5ms, for $H_\text{f0}=70\mu \text{m}$ at 3.1ms and for $H_\text{f0}=80\mu \text{m}$ no corona detachment is observed. It is apparent therefore, that the thickness of the corona liquid sheet at the instance of detachment is influenced by the thickness of the liquid film on the substrate. If this thickness is too large, no detachment of the corona occurs. 

In Figure~\ref{img:hc_exp_theory}  the predicted film thickness evolution according to Eq.~(\ref{eq:filmthicknesscrownwalldim}) and using a value of $B$ of 4 is also shown for the three substrate film thicknesses, as well be explained below. 

\section{Mechanism of the film disintegration by hole nucleation and expansion}
\subsection{Spontaneously growing holes in liquid films}
Examination of  images of disintegrating liquid films reveals that multiple holes often appear  prior to breakup  \citep[cf.][]{Brenn2005}.
These holes proliferate  into the intact film because of the surrounding circular free rims and are driven by  surface tension according to the Taylor-Culick mechanism \citep{Taylor1959,Culick1960,CollisionPhenomena2017}. These expanding holes then merge, leaving a network of ligaments, which break up due to capillary instability. This is the scenario observed and further explored analytically in the present work in relation to the mechanism of corona detachment.

Free liquid films, being a two-dimensional continuum, are not inherently prone to break up into droplets, because in contrast to one-dimensional continua (jets), their surface energy would increase through breakup. Therefore, holes  appear only as a result of a nucleation, which is, for example, a perturbaton resulting from disturbances in the corona wall film \citep{wakimoto2009influence}. 

%When a single disk-like  hole of  radius $r$ is formed in a film, its surface area, and thus surface energy, decrease by the value of

%where $\sigma$ is the surface tension  and the factor 2 accounts for the two surfaces of the film.

Consider a circular disk-like hole in a film of thickness $h$. The hole in the film is surrounded by a free rim of  cross-sectional radius $a$, which accommodates the liquid volume removed from the hole, i.e. $\pi (r+a)^2h$, where $r$ is the radius of the rim centerline. The rim volume is $2\pi^2ra^2$, and thus,  volume conservation yields
\begin{equation}
\label{eq:f2}
a=\frac{r h^{1/2}}{\sqrt{2\pi r}-h^{1/2}}
\end{equation}
 The surface energy increase attributed to the surface energy of the free rim, $\Delta\Phi_{rim}=4\sigma \pi^2 r a$ is expressed accounting for  Eq. (\ref{eq:f2}) as
\begin{equation}
\label{eq:f4}
\Delta\Phi_{rim}=\frac{4\sigma \pi^2 r^2 h^{1/2}}{\sqrt{2\pi r}-h^{1/2}},
\end{equation}
where  $\sigma$ is the surface tension. 

The surface energy decrease due to the hole formation is 
\begin{equation}
\label{eq:f1}
\Delta\Phi_{film}=-2\pi (r+a)^2 \sigma
\end{equation}
Here the factor 2 accounts for the two surfaces of the film. 

Equations (\ref{eq:f4}) and (\ref{eq:f1}) show with the help of (\ref{eq:f2}) that the total energy change related to the formation of a hole is

\begin{equation}
\label{eq:f5}
\Delta\Phi (r)=\Delta\Phi_{film}+\Delta\Phi_{rim}=-\frac{4 \pi ^2 r^2 \left(h-\sqrt{2 \pi r h } +r\right)
   \sigma }{\left(\sqrt{2 \pi r }-\sqrt{h} \right)^2}
\end{equation}
The function  $\Delta\Phi(r)$  has a maximum corresponding to the condition $d\Delta\Phi(r)/dr=0$ , which yields the critical hole nucleus size 

\begin{equation}
\label{eq:f6}
r_*\approx 2.18 h
\end{equation}
If the radius of a hole 'nucleus' $r$ is smaller than $r_*$, its growth would correspond to an increase in the total energy, given by Eq.~(\ref{eq:f5}), i.e., it is energetically unfavorable and thus, cannot be spontaneous. On the other hand, if a hole 'nucleus' is larger than the critical one, i.e., $r>r_*$  , its growth would correspond to a decrease in the total energy $\Delta\Phi(r)$  given by Eq.~(\ref{eq:f5}), i.e., it would be energetically favorable and thus, spontaneous. The critical total energy corresponding to the critical hole 'nucleus' is found as $\Delta\Phi_{*}=\Delta\Phi(r_*)$ , and according to Eqs.~(\ref{eq:f5}) and (\ref{eq:f6}), is equal to %\textcolor{teal}{$r$ or $r_*$?}
\begin{equation}
\label{eq:f7}
\Delta\Phi_* \approx 13.4 \sigma h^2
\end{equation}
Essentially, this is the activation energy required to be exceeded to form a spontaneously growing hole. 

Note that \cite{Taylor1973} developed an alternative approach to determine the size of a critical hole prone to grow. They argue that a hole in a liquid film in equilibrium with the surrounding vacuum should possess a catenoidal shape, which guarantees zero capillary pressure in the liquid on the hole banks in equilibrium with vacuum. There are two possible catenoidal hole shapes of different sizes, with the smaller one (in radius) being stable, and thus, non-growing, and the larger one being unstable, and thus, growing. The approach of \cite{Taylor1973} invokes the stability arguments based on the surface energy of catenoidal holes and yields $r_*\sim h$ [cf. Eq. (\ref{eq:f6})]. However, their approach implies hole formation process without formation of a free rim, which could be realized in the \cite{Taylor1973} apparatus with suspended soap films, but is not applicable to hole formation in dynamic free liquid films originating from drop impacts onto pre-existing liquid films (or many other films, e.g., those originating from swirl atomizers). Moreover, the requirement of zero capillary pressure in liquid in equilibrium with the surrounding vacuum is arbitrary. Indeed, a spherical drop is in equilibrium with the the surrounding vacuum with a non-zero capillary pressure. The present approach accounts for realistic configurations of holes with free rims, and moreover, does not insist that their growth inevitably begins from an  arbitrary chosen equilibrium catenoidal shape. 

\subsection{Turbulent eddies in liquid films}%as triggers of spontaneously growing holes in swirling films}

%The probability of a growing hole nucleus formation depends on the energy barrier described above and the energy level provided by perturbations (noise) $E_T$ (e.g., from the surrounding air flow, vibrations, etc.) Then, the probability P of the critical nucleus (a spontaneously growing hole) formation is found similarly to the Boltzmann distribution as
Consider a liquid film with disturbances or turbulent eddies inside it, as suggested by the experiments of \cite{wakimoto2009influence}. It should be emphasized that at the moment of drop impact onto a film on the wall and redirection of fluid flow along the wall, the Reynolds number is high enough to expect eddy formation. These eddies would also be entrained into the corona emerging from the film.  In the inertial range, down to the dissipation range, 
the distribution of pulsation energy by wavenumbers $\kappa$, $E(\kappa)$ is given by the Kolmogorov spectrum \citep{kolmogorov1962refinement,pope2001turbulent}:
\begin{equation}
\label{eq:pulsationenergy}    
E(\kappa)=C  \varepsilon^{2/3} \kappa^{-5/3} 
\end{equation}
where $\epsilon$ is the specific dissipation rate, and $C$ is the universal Kolmogorov spectrum  constant $C=1.5$ according to \cite{george1984pressure}, which is related to the experimentally determined spectral constant $C_k$ as $C=(55/18)C_K$ \citep{sreenivasan1995universality}

The specific pulsation energy $k$ of all the turbulent eddies in the film is found as

\begin{equation}
    \label{eq:specpulsationenergy}
    k=\int_{\kappa_h}^{\infty}E(\kappa')dk'=\frac{3}{2}C\biggl(\frac{\varepsilon h}{2\pi}\biggr)^{2/3}
\end{equation}

\noindent where $\kappa_h=2\pi/h$, and $\kappa'$ is the dummy variable. 

The Kolmogorov length scale is $\eta=(\nu^3/\varepsilon)^{1/4}$, with $\nu$ being the kinematic viscosity. Assuming the overlap of the inertial and dissipation ranges, take $\eta \approx h$. Then, the dissipation rate is estimated as

\begin{equation}
    \label{eq:dissipationrate}
    \varepsilon \approx \frac{\nu^3}{h^4}
\end{equation}

\noindent Using Eqs.~(\ref{eq:specpulsationenergy}) and (\ref{eq:dissipationrate}), one finds the kinetic energy of turbulence per unit volume, $E_T=\rho k$, as 
\begin{equation}
\label{eq:KinEnTurb}
 E_T=\frac{3C}{2(2\pi)^{2/3}} \frac{\rho \nu^2}{h^2}
\end{equation}
The latter shows that the specific pulsation energy increases in smaller eddies (thinner film), as in the Kolmogorov theory. 

Consider the film as a system of turbulent eddies, and introduce its temperature in the energy units $T_e$ as
\begin{equation}
    \label{eq:Temperature}
    T_e=E_T 
\end{equation}
 For the system of turbulent eddies one can, essentially, repeat the entire thermodynamical derivation starting from the microcanonic $\delta$-functional distribution to the introduction of the entropy $S$ as in \cite{landau2013statistical}

\begin{equation}
\label{eq:entropy}
    \frac{dS}{dE}=\frac{1}{T_e}
\end{equation}
where $E$ is understood here as the turbulence energy.
 Then, the probability of a system of ‘turbulent eddies and a critical hole’ is given by the Gibbs distribution \citep{landau2013statistical}
\begin{equation}
\label{eq:GibbsDist}
    P=K\text{exp}\biggl(-\frac{\Delta\Phi_*'}{T_e}\biggr)
\end{equation}

\noindent which is also called the Boltzmann distribution; $K$ is a dimensionless constant (not to be confused with the convention of using $K$ in Eqs.~(\ref{eqCoss}) - (\ref{eq:K}).

In Eq.~(\ref{eq:GibbsDist}), $\Delta \Phi_*'=\Delta \Phi/\pi[ r_* + a(r_*)]^2 h$ is the hole energy per unit volume.
%(\textcolor{red}{Ilia: here you can also do $(r_*+a(r_*))^2$, as you did before, which will be beneficial for a better comparison with the theory}) 
According to Eqs.~(\ref{eq:f6}) and (\ref{eq:f7}), $\Delta \Phi_*'\approx 0.48 \sigma/h$. Then, Eqs.~(\ref{eq:KinEnTurb}), (\ref{eq:Temperature}) and (\ref{eq:GibbsDist}) yield

\begin{equation}
    \label{eq:PropabilityK}
    P \approx K\text{exp}\left[- \frac{1.09}{C}\frac{\sigma h}{\rho \nu^2}\right].
\end{equation}
 At $h=0$, the probability of a system with a critical hole is 1, which yields $K=1$, and thus,
\begin{equation}
    \label{eq:Propability}
    P=\text{exp}\left[- \frac{1.09}{C}\frac{\sigma h}{\rho \nu^2}\right].
\end{equation}
Equation~(\ref{eq:Propability}) shows that the thinner  the film is, the higher is the probability of a critical hole forming. Taking for an estimate the parameters of water, $\rho = 1 \thinspace \text{g/cm}^3$, $\sigma=72 \thinspace \text{g/s}^2,  \nu=10^{-2} \thinspace\text{cm}^2/\text{s}$, and the above-mentioned value of the empirical constant $C=1.5$, one obtains $P=0.59$ for $h=10$ nm. The estimation for the silicon oil S10 yields $P=0.49$ ($\rho$=0.93 g/cm$^3$)  for $h=5$ $\mu$m.  With a lower value of the empirical constant $C$, the probability of critical hole formation in thicker films becomes higher. 

%\begin{equation}
%\label{eq:3f4}
%P=\text{exp}\Bigg(-\frac{\Delta\Phi_{*}}{E_T}\Bigg)
%\end{equation}
%For some realistic estimates of $E_T$, one finds
%\begin{equation}
%\label{eq:3f6}
%P\sim \begin{cases}
%0.164     & \text{ for } h=10^{-2}\text{cm} \\
%0  & \text{ for } h=10^{-1}\text{cm} 

%\end{cases}
%\end{equation}

%\noindent which shows that formation of spontaneously growing holes becomes more  probable as the film thickness decreases.
\subsection{Evolution of film thickness in time}
The probability of critical hole formation (\ref{eq:Propability}) depends on time because the corona wall thickness $h$ depends on time. Consider the simplest case where the pre-existing film on the wall and the impacting drop are of the same liquid. In this case the experiments show that corona detachment is possible. Then, the theory of \cite{yarin1995impact} is applicable and the dimensionless radial velocity in the film in the wall at the spreading corona,   $\tilde{U}_c$, is given by 
\begin{equation}
\label{eq:velradfilm}
\tilde{U}_c=\frac{B \tilde{r}_c(\tilde{t}-\tilde{t}_0)}{1+B(\tilde{t}-\tilde{t}_0)}
\end{equation}
where $B$ is dimensionless and determined by the radial velocity gradient in the film on the wall resulting from the drop impact near the impact center. Note that here and hereinafter time is rendered dimensionless using $D_0/U_0$, velocity using $U_0$ and lengths using $D_0$. Dimensionless times, lengths and velocities are signified with a tilde. The time shift $\tilde{t}_0$ is required because, as discussed in \cite{yarin1995impact}, this theory describes only a remote asymptotics and cannot be extended to the drop impact time $\tilde{t}=0$; cf. $\tilde{\tau}$ in Eq.(\ref{eq:yarin}). The shift is equivalent to the 'polar distance' introduced when the theory of self-similar submerged jets, valid as remote asymptotics, is compared to the experimental data acquired using jets issued from a finite nozzle. \\

\noindent In addition,
\begin{equation}
    \label{eq:radCrownPos}
 \tilde{r}_c=\sqrt{2A(\tilde{t}-\tilde{t}_0)}
\end{equation}
is the current dimensionless radial position of the corona, with  $A$ being the dimensionless integral characteristic of the radial velocity distribution in the film on the wall resulting from the drop impact. Accordingly,

\begin{equation}
    \label{eq:velradfilmCrownPos}
    \tilde{U}_c=B\sqrt{2 A}\frac{\sqrt{\tilde{t}-\tilde{t}_0}}{1+B(\tilde{t}-\tilde{t}_0)}
\end{equation}
Assuming negligible viscous losses when the liquid is propelled from the film to the corona, one can find the dimensionless corona height $\tilde{L}$ integrating the following equation: 

\begin{equation}
    \label{eq:diffcrownheight}
    \frac{d\tilde{L}}{d\tilde{t}}=\tilde{U}_c
\end{equation}
which together with Eq.~(\ref{eq:velradfilmCrownPos}) yields
\begin{equation}
\label{eq:crownheight}
    \tilde{L}=2\sqrt{2A}\left[ \sqrt{\tilde{t}-\tilde{t}_0}-\frac{1}{\sqrt{B}}\text{arctan}\sqrt{B(\tilde{t}-\tilde{t}_0)}\right].
\end{equation}
For relatively short times of interest here, Eq.~(\ref{eq:crownheight}) yields
\begin{equation}
\label{eq:crownheightshortt}
    \tilde{L}=\frac{2\sqrt{2A}B}{3}(\tilde{t}-\tilde{t}_0)^{3/2}
\end{equation}
Accordingly, the current volume of the corona is
\begin{equation}
\label{eq:Volcrown}
   \tilde{V}_{\textrm{corona}}= 2\pi \tilde{r}_c \tilde{h} \tilde{L} 
\end{equation}
whereby the volume is rendered dimensionless with $D_0^3$. On the other hand, this volume was propelled from the film on the wall inside the corona, i.e., 
\begin{equation}
\label{eq:Volcrownhf}
    \tilde{V}_{\textrm{corona}}=\pi \tilde{r}_c^2(\tilde{\delta}_\mathrm{f0}-\tilde{H}_f) , \quad \tilde{\delta}_\mathrm{f0} = \frac{H_\mathrm{f0}}{D_0}
\end{equation}
where $\tilde{H}_f$ is the current dimensionless film thickness of the wall inside the corona. 
%initial dimensional thickness of the film on the wall, $\delta_{\textrm{f0}}$ is its  dimensionless value and $D_0$ is the initial drop diameter. 

The film thickness rendered dimensionless by $D_0$ is found as  \citep{yarin1995impact}
\begin{equation}
\label{eq:filmthicknessh_f}
   \tilde{H}_f=\frac{\tilde{\delta}_{\textrm{f0}}}{[1+B(\tilde{t}-\tilde{t}_0)^2]}
\end{equation}
Then, using Eqs.~(\ref{eq:radCrownPos}) and (\ref{eq:crownheightshortt}) - (\ref{eq:filmthicknessh_f}) one obtains the dimensionless thickness of the corona wall as
\begin{equation}
\label{eq:filmthicknesscrownwall}
    \tilde{h}_c=\frac{3 \tilde{\delta}_{\textrm{f0}}}{4 B (\tilde{t}-\tilde{t}_0)} \left\{1-\frac{1}{[1+B(\tilde{t}-\tilde{t}_0)^2]}\right\}.
\end{equation}
In  dimensional form, Eq.~(\ref{eq:filmthicknesscrownwall}) reads
\begin{equation}
    \label{eq:filmthicknesscrownwalldim}
    h_c=H_\text{f0}\frac{3D_0}{4B U_0 (t-t_0)}\left\{1-\frac{1}{1+B [U_0(t-t_0)/D_0]^2}\right\}
\end{equation}
%where $U_0$ is the impact velocity.
As $(t-t_0) \xrightarrow[]{} 0$, Eq.~(\ref{eq:filmthicknesscrownwalldim}) yields
\begin{equation}
      \label{eq:limit_value}
    h_c=\frac{3}{2}H_\text{f0}
\end{equation}
Then, according to Eq.~(\ref{eq:filmthicknesscrownwalldim}), $h_c$ decreases monotonically in time approximately as
\begin{equation}
\label{eq:hdecrease}
h_c=H_\text{f0}\frac{3D_0}{4B U_0 (t-t_0)}    
\end{equation}
According to Figure~\ref{img:hc_exp_theory}, the corona sheet would break up at 

\begin{equation}
\label{eq:hfrozen}
    h_c=h_b \approx 1 \mu\textrm{m}
\end{equation}
and a reasonable estimate of the  probability of hole formation would be, according to Eqs.~(\ref{eq:Propability}) , 
\begin{equation}
    \label{eq:frozenprobability}
    P=\text{exp}\left[-\frac{1.09}{C}\frac{\sigma h_b}{\rho\nu^2}\right]
\end{equation}

\subsection{Hole growth process and the corona detachment time}
Toroidal free rims surrounding the super-critical, spontaneously growing holes move outward with the Taylor-Culick velocity \citep{Taylor1959,Culick1960,CollisionPhenomena2017},
\begin{equation}
\label{eq:f8}
u_{\textrm{TC}}=\sqrt{\frac{2\sigma}{\rho h}}
\end{equation}

Consider a specific location in the film and hole forming around this location. Any hole which appears at a distance $R$ from the location at time $\tau < t - R/u_{\textrm{TC}}$ will reach this location before the instant $t$. The number of the holes $\mathrm{d} n$, formed in a ring of the radius $R$ defined by the radius element $\mathrm{d}R$, which reach the considered point at time $t$ can be estimated as
\begin{equation}
\label{eq:f9}
\mathrm{d} n=\int_{0}^{t-R/u_\textrm{TC}} I_* 2 \pi R \mathrm{d}R \mathrm{d}\tau=I_* \left( t-\frac{R}{u_{\textrm{TC}}}\right) 2\pi R \mathrm{d}R
\end{equation} 
where $\tau$ is an integration time, and $I_*$ is the constant  hole formation rate. 

Note that if  variation of $I_*$ in time would be accounted for, one would have to evaluate the integral $\int_{0}^{t-R/u_{\text{TC}}}I_* d\tau$ numerically, which would complicate the following calculations, but, in principle, does not affect the main theoretical structure.

In total, the average number of holes $N$ which can reach the location under consideration during time $t$ is  
\begin{equation}
\label{eq:f10}
N=\int_{0}^{u_{TC}t}I_*\left( t-\frac{R}{u_{\textrm{TC}}}\right) 2\pi R dR=I_* \frac{\pi}{3}\text{u}_{\textrm{TC}}^2 t^3
\end{equation}
Because the growing hole formation process is random, the probability that the location under consideration will be reached by $m$ holes during time $t$ is given by the Poisson distribution
\begin{equation}
\label{eq:f11}
P_m(t)=\frac{N^m}{m!}\text{exp}(-N)
\end{equation}
Then, the probability that zero holes will reach that location $(m=0)$, i.e., it will stay intact is equal to 
\begin{equation}
\label{eq:f12}
P_0(t)=\text{exp}(-N)
\end{equation}
Accordingly, the relative area occupied by the holes, accounting for their interactions, is
\begin{equation}
\label{eq:f13}
\lambda=1-\text{exp}(-N)
\end{equation}

%
%
%This shows that
%\begin{equation}
%\label{eq:f14}
%\frac{d\eta}{dt}=(1-\eta)\frac{dN}{dt}
%\end{equation}
%where d/dt denotes the material time differentiation.
%According to Eq.~(\ref{eq:f10}),
%\begin{equation}
%\label{eq:f15}
%\frac{dN}{dt}=I_*\pi \text{v}_{TC}^2t^2
%\end{equation}
%Also, the number of critical holes
%resulting in growing holes which appear per unit area
%\begin{equation}
%\label{eq:f16}
%c=\int_{0}^{t}I_*\,d\tau=I_* t
%\end{equation}
%(the units of $c$ are $1/\text{cm}^2$.) Then, Eq. (\ref{eq:f15}) takes the form
%
%\begin{equation}
%\label{eq:f17}
%\frac{dN}{dt}=c\pi \text{v}_{TC}^2t
%\end{equation}
%
%Using Eqs.~(\ref{eq:f14}) and (\ref{eq:f17}), as well as Eq.~(\ref{eq:f16}), one obtains the system of equations describing the hole formation and growth mechanism and the film disappearance
%
%\begin{equation}
%\label{eq:f18}
%\frac{dc}{dt}=I_*
%\end{equation}
%
%\begin{equation}
%\label{eq:f19}
%\frac{d\eta}{dt}=(1-\eta)c\pi \text{v}_{\textrm{TC}}^2 t
%\end{equation}
%They are solved with the initial conditions
%\begin{equation}
%\label{eq:f20}
%t=0, \eta=c=0
%\end{equation} 
\noindent Note that the calculation of $\lambda$ via Eqs.~(\ref{eq:f9})-(\ref{eq:f13}) is similar in a sense to the calculation of the degree of crystallization in polymer crystallization processes \citep{yarin1992flow,yarin1993free,ghosal2019modeling}.

The characteristic time scale of Kolmogorov’s eddies is $\tau_\eta=(\nu/\varepsilon)^{1/2}$ (Kolmogorov 1962, Pope 2000). Using Eq.~(\ref{eq:dissipationrate}), one obtains
\begin{equation}
    \label{eq:KolTime}
    \tau_\eta=\frac{h^2}{\nu}
\end{equation}
Then, the specific rate of formation of growing holes per surface area is 
\begin{equation}
\label{eq:rateofformationtau}
    I_*=\frac{P}{(\pi r_*^2)(k \tau_\eta)}
\end{equation}
\noindent where $k$ is the number of characteristic time scales required for a hole formation.
Using Eqs.~(\ref{eq:f6}), (\ref{eq:KolTime}) and (\ref{eq:rateofformationtau}), one obtains
\begin{equation}
    \label{eq:rateofformationnuh}
 I_*=0.067\frac{\nu}{k h^4}P
\end{equation}
The  value of $I_*$ found from Eqs.~(\ref{eq:limit_value}), (\ref{eq:frozenprobability}) and (\ref{eq:rateofformationnuh}) reads
\begin{equation}
\label{eq:frozenrateofformation}
       I_*=0.013 \frac{\nu}{k H_\text{f0}^4}\text{exp}\left(-\frac{1.09}{C}\frac{\sigma h_b}{\rho \nu^2}\right)
\end{equation}

The expression Eqs.~(\ref{eq:f13}) for the relative area of the holes $\lambda$  with the help of Eqs.~(\ref{eq:limit_value}), (\ref{eq:f8}) and (\ref{eq:f10}) yields
\begin{equation}
    \label{eq:intpercolationeta}
    \lambda=1-\text{exp}\left(-\frac{4 \pi I_* \sigma }{9 \rho H_\text{f0}} t^3 \right)
\end{equation}

According to  percolation theory \citep{Stauffer1979,Stauffer1985}, when the value of 
\begin{equation}
\label{eq:f22}
\lambda=\frac{1}{2}
\end{equation}
has been reached, the intact film disappears.
Then, Eqs.~(\ref{eq:intpercolationeta}) and (\ref{eq:f22}) yield the corona detachment time $t_\text{d}$ as 
\begin{equation}
    \label{eq:crowndetachmenttime}
    t_\text{d}=\left(\frac{9\text{ln}2}{4\pi}\frac{\rho H_\text{f0}}{I_* \sigma}\right)^{1/3}
\end{equation}
According to Eqs.~(\ref{eq:frozenrateofformation}) and (\ref{eq:crowndetachmenttime}) the dependence of the corona detachment time $t_\text{d}$ on the initial film thickness on the wall $H_\text{f0}$ is 
\begin{equation}
     \label{eq:detachmenttimefilmthickness}
     t_\text{d}=3.37 \left(\frac{\rho k H_\text{f0}^5}{\sigma \nu}\right)^{1/3}  \text{exp}\left(\frac{0.363}{C}\frac{\sigma h_b}{\rho \nu^2}\right)
\end{equation}
using Eq.~(\ref{eq:hfrozen}) the exponent in Eq.~(\ref{eq:detachmenttimefilmthickness}) with $h_b \approx$ 1 $\mu$m and $C$=1.5 is approximately 1.  Then,  Eq.~(\ref{eq:detachmenttimefilmthickness}) yields  
\begin{equation}
    \label{eq:detachmentmenttimefilmthicknessexp1}
    t_\text{d} \approx 3.37 k^{1/3}\left(\frac{\rho H_\text{f0}^5}{\sigma \nu}\right)^{1/3} 
\end{equation}
Equation~(\ref{eq:detachmentmenttimefilmthicknessexp1}) predicts a  scaling $t_d \sim H_\text{f0}^{5/3}$, which has been added to the log-log representation of the experimental data in Figure~\ref{img:S2}. The exact position of this scaling relation on the diagram was chosen using a least squares fit to the data, whereby  $k$ was used as a fitting parameter.    The comparison shown in Figure~\ref{img:S2} indicates that this scaling describes the experimental results very well for corona detachments with like fluids. The experimental data suggests that this slope may increase slightly with decreasing value of $\tilde{\kappa}_{\nu}$, i.e., when the film liquid becomes less viscous than the drop liquid. 

\subsection{Detachment of the corona formed by droplet impact onto a layer of another liquid}

When a drop impacts onto a liquid layer, it transforms into a disk practically without viscous losses \cite{yarin1995impact}. This transformation takes time of the order of $D_0/U_0$. Because viscous losses are neglected, the radial velocity of spreading is still of the order of $U_0$.
Accordingly, the resulting disk radius is equal to $D_0$. Then, the initial disk thickness $H_\text{fd0}$ is found from the  volume conservation condition
$\pi D_0^2 H_\text{fd0}=\pi D_0^3/6$ as
\begin{equation}
    \label{eq:diskheight}
    H_\text{fd0}=\frac{D_0}{6}
\end{equation}
Taking for the estimate $D_0 \sim 10^{-1}\text{cm}$, one finds $H_\text{fd0}=170~\mu\text{m}$, which is comparable or thicker than many pre-existing liquid films on the wall. Because of the appropriate profile of the radial velocity in this disk-like film, corona formation will be driven by it, and liquid radially outflowing from the impact point will be propelled into the corona.
However, if (and only if!) a different liquid is contained in the pre-existing liquid film, then, the velocity and shear stresses should be continuous at the liquid-liquid interface, which yields the following condition
\begin{equation}
\label{eq:shearstressbalance}
    \mu_1=\frac{U_{1\text{s}}}{\delta_1}=\mu_2\frac{U_{2\text{s}}-U_{1\text{s}}}{\delta_2}
\end{equation}
where subscript 1 corresponds to the pre-existing liquid film, and subscript 2 corresponds to the film formed by the drop liquid. Also, the dynamic viscosities of the liquids are denoted as $\mu_1$ and $\mu_2$, the dimensional interfacial velocity is denoted as $U_{1\text{s}}$, and the velocity at the free surface  as $U_{2\text{s}}$. In addition, the dimensional viscous lengths are denoted as $\delta_1$ and $\delta_2$.
It should be emphasized that Eq.~(\ref{eq:shearstressbalance}) is invalid if the liquids in the drop and the pre-existing liquid film are the same.

Note that approximately
\begin{equation}
\label{eq:viscouslength}
\delta_1=\sqrt{\nu_1 (t-t_0)}, \quad \delta_2=\sqrt{\nu_2 (t-t_0)}
\end{equation}
where $\nu_1$ and $\nu_2$ are the kinematic viscosities, and $t$ is dimensional time. 
Accordingly, 
\begin{equation}
    \label{eq:interfacialvelocities}
    U_{1\text{s}}=\frac{s}{1+s}U_{2\text{s}}, \quad s=\sqrt{\frac{\mu_2}{\mu_1}\frac{\rho_2}{\rho_1}}
\end{equation}
where $\rho_1$ and $\rho_2$ are liquid densities, and the velocities can either still be dimensional, or rendered dimensionless by the same velocity scale.

Consider entrainment of liquid from the pre-existing film on the wall. The radial velocity profile in it can be taken in the first approximation as
\begin{equation}
\label{eq:radialvelprof}
    v_r=V_{1\text{s}}\frac{\tilde{y}}{\tilde{\delta}_1}
\end{equation}
where $\tilde{y}$ is the normal coordinate reckoned from the wall (where $\tilde{y}$=0), and $\tilde{y}$ and $\tilde{\delta}_1$ are rendered dimensionless by $H_\text{f0}$ and velocities with $U_0$, as in section 4. Accordingly, in  dimensionless form the first expression from Eq.~(\ref{eq:viscouslength}) reads
\begin{equation}
    \tilde{\delta}_1=M\sqrt{\tilde{t}-\tilde{t}_0},\quad M=\sqrt{\frac{\nu_1 D_0}{U_0 H_{\text{f0}}^2}}
\end{equation}
using the same scale for time as in section 4 or \cite{yarin1995impact}. Then, at the corona, $\tilde{U}_{2\text{s}}=\tilde{U}_c$, with the latter being given by Eq.~(\ref{eq:velradfilmCrownPos}), i.e.,
\begin{equation}
\label{eq:interfacialvel2}
    \tilde{U}_{2\text{s}}=B\sqrt{2A}\frac{\sqrt{\tilde{t}-\tilde{t}_0}}{1+B(\tilde{t}-\tilde{t}_0)}
\end{equation}
The dimensionless volumetric flow rate of liquid from the pre-existing liquid film entrained into the radial motion and ultimately propelled into the corona is found as
\begin{equation}
 \tilde{\Dot{V}}=\int_0^{\tilde{\delta}_1}\tilde{v}_r\,d\tilde{y}2\pi \tilde{r}_c
\end{equation}
The volumetric flow rate is rendered dimensionless by $U_0D_0H_{\textrm{f0}}$.
Using Eqs.~(\ref{eq:interfacialvelocities})-(\ref{eq:interfacialvel2}) and (\ref{eq:radCrownPos}), the latter yields
\begin{equation}
\label{eq:Volrate2}
    \tilde{\Dot{V}}=\frac{s}{s+1}(2\pi ABM)\frac{(\tilde{t}-\tilde{t}_0)^{3/2}}{1+B(\tilde{t}-\tilde{t}_0)}
\end{equation}
Using Eq.~(\ref{eq:Volrate2}), one finds the dimensionless volume of the pre-existing liquid film entrained into the crown by time $\tilde{t}$ as
\begin{equation}
\label{eq:Volent}
\begin{split}
    \tilde{V}_\text{ent} &=\frac{s}{s+1}(2\pi ABM) \int_{0}^{(\tilde{t}-\tilde{t}_0)}\frac{\tilde{t}^{3/2}}{1+B\tilde{t}}d\tilde{t} \\[10pt]
    &=\frac{s}{s+1}(2\pi ABM)\left\{\left[\frac{2(\tilde{t}-\tilde{t}_0)^{3/2}}{3B}-\frac{2(\tilde{t}-\tilde{t}_0)^{1/2}}{B^2} \right]+\frac{2}{B^{5/2}}\text{arctan}\sqrt{B(\tilde{t}-\tilde{t}_0)}\right\}
\end{split}
\end{equation}
This volume is rendered dimensionless by $D_0^2H_{\textrm{f0}}$. In the short-time limit of interest here, Eq.~(\ref{eq:Volent}) yields
\begin{equation}
\label{eq:Volentshorttime}
    \tilde{V}_\text{ent}=\frac{4}{5}\frac{s}{(s+1)}\pi ABM(\tilde{t}-\tilde{t}_0)^{5/2}
\end{equation}
The dimensionless thickness of the corona wall formed by liquid 1 entrained from the pre-existing film at the wall, $h_1$, is found from the condition employing Eq.~(\ref{eq:Volentshorttime}) as
\begin{equation}
\label{eq:preh1}
    2\pi \tilde{r}_c \tilde{L} \tilde{h}_1=\frac{4}{5}\frac{s}{(s+1)}\pi ABM(\tilde{t}-\tilde{t}_0)^{5/2}
\end{equation}
The thickness $h_1$ is rendered dimensionless by $H_{\textrm{f0}}$. Using Eqs.~(\ref{eq:radCrownPos}) and (\ref{eq:crownheightshortt}), one finds from Eq.~(\ref{eq:preh1})
\begin{equation}
\label{eq:wallfilmfluidcrownshare}
\tilde{h}_1=\frac{3}{10}\frac{s}{s+1}M(\tilde{t}-\tilde{t}_0)^{1/2}
\end{equation}
The corresponding dimensional expression takes the form
\begin{equation}
\label{eq:wallfilmcrownsharedim}
    h_1=\frac{3}{10}\frac{\sqrt{\mu_2 \rho_2 / \rho_1^2}}{\sqrt{(\mu_2/\mu_1)(\rho_2/\rho_1)}+1}(t-t_0)^{1/2}
\end{equation}
The latter shows that if the droplet liquid is much more viscous than the liquid in the pre-existing liquid film on the wall, i.e., $\mu_2 /\mu_1 >>1$, then
\begin{equation}
    \label{eq:h1viscousdrop}
    h_1=\frac{3}{10}\sqrt{\frac{\mu_1}{\rho_1}}(t-t_0)^{1/2}
\end{equation}
In the opposite limit, $\mu_2/\mu_1 <<1$, where the droplet liquid is much less viscous than the liquid in the pre-existing film on the wall, Eq.~(\ref{eq:wallfilmcrownsharedim}) yields
\begin{equation}
\label{eq:h1viscousfilm}
    h_1=\frac{3}{10}\sqrt{\mu_2\rho_2}\frac{1}{\rho_1}(t-t_0)^{1/2}
\end{equation}
 
Equations~(\ref{eq:h1viscousdrop}) and (\ref{eq:h1viscousfilm}) express therefore the part of the corona wall thickness comprised of fluid from the pre-existing wall film  depending on the ratio of dynamic viscosities. In both cases the thickness increases with the same time dependence. This theoretical result is not straightforward to confirm experimentally, but is the subject of further investigation.

%\textcolor{red}{there must be a better connection between Alex's part on hole formation and on the hole formation observed in experiments.}\textcolor{teal}{We have to compute these equations. But need to estimate evolution of $h$ in time. For computations we also need estimations for $K$ and for $E_T$. To coordinate with Yarin. It would be excellent to estimate $K$ and $E_T$ from another problem/paper and to apply them here. For example from the results of Opfer. We have his times for the hole formation. }

\section{Conclusions}
In this study the impact of a liquid drop onto a solid substrate pre-wetted by a film of the same or another liquid is studied, and a fascinating phenomenon of a complete, almost instantaneous detachment of the cylindrical corona sheet from the pre-existing liquid film on the wall is observed. The viscosity and the viscosity ratio   of these liquids is varied over a wide range, through which conditions for a detachment of the corona liquid sheet from the wall film could be experimentally investigated. 
Focus was then placed on the physics behind the almost instantaneous and uniform corona detachment, since the breakup of liquid sheets is a fundamental step in numerous atomizers, such as pressure swirl, flat fan nozzles, or as a limiting factor in thin-film coating processes.

Even though liquid is permanently propelled into corona sheet and gravity plays practically no role, the thinning of the corona sheet with time is predicted theoretically, exhibiting excellent
agreement with experimental findings. This is related to the fact that the corona radius expansion in time appears to be the dominant process, which determines the corona sheet thinning. For this comparison between the theory and experiment, the Taylor-Culick relation was employed in an innovative manner as a means for measuring the local, instantaneous thickness of the corona sheet at or near the time of rupture. This was made possible by the use of high-speed videos, which could capture the movement of the free rim at the perimeter of the rupture holes in the corona sheet. This method can easily be applied as a film thickness measurement of other rupturing liquid films.

It is shown that an ongoing thinning of the corona sheet with time makes the  nucleation of ‘super-critical’ (by radius) holes more probable, i.e. those holes whose growth in time is sustained by a decrease in the total surface energy in the system ‘corona sheet with a hole surrounded by a free rim’. Accordingly, growth of super-critical holes is shown to be energetically favorable. The hole nuclei are attributed to Kolmogorov eddies resulting from the turbulent eddy cascade generated by a strong shear field at the interface of the impacting drop and the pre-existing liquid film on the wall. Being entrained in the corona sheet, turbulent eddies rupture the hole nuclei with a high probability if the sheet is sufficiently thin ($\sim$ 1 $\mu$m). That explains why a ‘perfect corona detachment’ (with a sharp uniform cut-off) happens at its base, where the eddies are entrained,  resulting almost instantaneously in  hole nuclei, which rapidly increase in size and merge. The probability of a system of ‘turbulent eddies and a critical hole’ is given by the Gibbs distribution rooted in the microcanonic $\delta$-functional distribution of thermodynamics.

Moreover, the growth process of the super-critical holes in time is described using the Taylor-Culick formula for the velocity of propagation of a free rim over a corona sheet, and thus the time required for super-critical holes to break the intact corona sheet up is predicted in the framework of the percolation theory. This is the time of corona detachment $t_d$, and it is shown theoretically that this time is related to the thickness of the pre-existing liquid film on the wall $H_\textrm{f0}$ by the following scaling law: $t_d \sim H_\textrm{f0}^{5/3}$. This law is in excellent agreement with the experimental findings in the case of the same liquids in the impacting drop and the pre-existing liquid film, i.e., in the case for which this theory has been derived. In the case of dissimilar liquids in the impacting drop and the pre-existing liquid film on the wall, the experimental data show that the scaling apparently slightly changes depending on the viscosity ratio.

The case of dissimilar liquids is also studied theoretically and the thickness of the part of the corona sheet comprised of the liquid from the pre-existing liquid film on the wall is predicted as a function of time and the viscosities and densities of both liquids (in the pre-existing film and the impacting drop). This theoretical result is not currently straightforward to verify experimentally, but can probably be tackled in future research.

In the broader context a ‘perfect corona detachment’ discovered and explained in this work is a manifestation of the almost instantaneous rupture processes characteristic of liquid and solid bodies with high ‘frozen-in’ skin or body stresses. In the present case this is a very thin corona sheet, which is almost instantaneously cut off from the pre-existing liquid film at the wall by the ‘frozen-in’ surface tension, which rapidly increases super-critical holes nucleated by entrained turbulent eddies.  A solid-state analog is  provided by such glass items as Prince Rupert's drops (aka Dutch or Batavian tears) which possess high ‘frozen-in’ residual stresses resulting from the production and reveal an explosive disintegration if the tail end is even slightly damaged. Another example of this type is given by Scirpus plants (reed, club-rush, wood club-rush or bulrush) whose ripe ‘flowers’ on top of a dry cane are brown cylindrical tightly packed clusters of small spikelets. After being slightly touched, the spikelets release themselves from the pack practically instantaneously in a sparkling champagne-like way and immediately become airborne.

\section*{Acknowledgement}
This research was supported by the the German Research Foundation (Deutsche Forschungsgemeinschaft) in the framework of the SFB-TRR 150 (Project ID: 237267381) collaborative research center, subproject A02.
\bibliographystyle{jfm}
\bibliography{ArxivCoronaDetachment}

\begin{thebibliography}{44}
\expandafter\ifx\csname natexlab\endcsname\relax\def\natexlab#1{#1}\fi
\def\au#1{#1} \def\ed#1{#1} \def\yr#1{#1}\def\at#1{#1}\def\jt#1{\textit{#1}}
  \def\bt#1{#1}\def\bvol#1{\textbf{#1}} \def\vol#1{#1} \def\pg#1{#1}
  \def\publ#1{#1}\def\arxiv#1{#1}\def\org#1{#1}\def\st#1{\textit{#1}}

\bibitem[Bakshi {\em et~al.\/}(2007)Bakshi, Roisman \&
  Tropea]{bakshi2007investigations}
{\sc \au{Bakshi, S.}, \au{Roisman, I.~V.} \& \au{Tropea, C.}} \yr{2007}
  \at{Investigations on the impact of a drop onto a small spherical target}.
  \jt{Phys. Fluids}  \bvol{19}~(3),  \pg{032102}.

\bibitem[Bolleddula {\em et~al.\/}(2010)Bolleddula, Berchielli \&
  Aliseda]{bolleddula2010impact}
{\sc \au{Bolleddula, D.~A.}, \au{Berchielli, A.} \& \au{Aliseda, A.}} \yr{2010}
   \at{Impact of a heterogeneous liquid droplet on a dry surface: Application
  to the pharmaceutical industry}.  \jt{Adv. Colloid Interface Sci.}
  \bvol{159}~(2),  \pg{144--159}.

\bibitem[Brenn {\em et~al.\/}(2005)Brenn, Prebeg, Rensink \& Yarin]{Brenn2005}
{\sc \au{Brenn, G.}, \au{Prebeg, Z.}, \au{Rensink, D.} \& \au{Yarin, A.~L.}}
  \yr{2005}  \at{The control of spray formation by vibrational excitation of
  flat-fan and conical liquid sheets}.  \jt{Atomization and Sprays}
  \bvol{15}~(6),  \pg{661--685}.

\bibitem[Chiu \& Lin(2005)]{chiu2005experiment}
{\sc \au{Chiu, S.-L.} \& \au{Lin, T.-H.}} \yr{2005}  \at{Experiment on the
  dynamics of a compound drop impinging on a hot surface}.  \jt{Phys. Fluids}
  \bvol{17}~(12),  \pg{122103}.

\bibitem[Cossali {\em et~al.\/}(1997)Cossali, Coghe \&
  Marengo]{cossali1997impact}
{\sc \au{Cossali, G.~E.}, \au{Coghe, A.} \& \au{Marengo, M.}} \yr{1997}
  \at{The impact of a single drop on a wetted solid surface}.  \jt{Exp. Fluids}
   \bvol{22}~(6),  \pg{463--472}.

\bibitem[Culick(1960)]{Culick1960}
{\sc \au{Culick, F.}} \yr{1960}  \at{Comments on a ruptured soap film}.  \jt{J.
  Appl Phys.}  \bvol{31},  \pg{1128--1129}.

\bibitem[Derby(2010)]{derby2010inkjet}
{\sc \au{Derby, B.}} \yr{2010}  \at{Inkjet printing of functional and
  structural materials: fluid property requirements, feature stability, and
  resolution}.  \jt{Ann. Rev. Mater. Res.}  \bvol{40},  \pg{395--414}.

\bibitem[Derby \& Reis(2003)]{derby2003inkjet}
{\sc \au{Derby, B.} \& \au{Reis, N.}} \yr{2003}  \at{Inkjet printing of highly
  loaded particulate suspensions}.  \jt{MRS Bull.}  \bvol{28}~(11),
  \pg{815--818}.

\bibitem[George {\em et~al.\/}(1984)George, Beuther \&
  Arndt]{george1984pressure}
{\sc \au{George, W.~K.}, \au{Beuther, P.~D.} \& \au{Arndt, R. E.~A.}} \yr{1984}
   \at{Pressure spectra in turbulent free shear flows}.  \jt{J. Fluid Mech}
  \bvol{148},  \pg{155--191}.

\bibitem[Geppert {\em et~al.\/}(2016)Geppert, Terzis, Lamanna, Marengo \&
  Weigand]{LamannaGeppert2016}
{\sc \au{Geppert, A.}, \au{Terzis, A.}, \au{Lamanna, G.}, \au{Marengo, M.} \&
  \au{Weigand, B.}} \yr{2016} Two component droplet wall-film interactions:
  Impact dynamics on very thin films.  \bt{In {\em Proc. ILASS-Europe 2016\/}}.
  Brighton, UK.

\bibitem[Ghosal {\em et~al.\/}(2019)Ghosal, Chen, Sinha-Ray, Yarin \&
  Pourdeyhimi]{ghosal2019modeling}
{\sc \au{Ghosal, A.}, \au{Chen, K.}, \au{Sinha-Ray, S.}, \au{Yarin, A.~L.} \&
  \au{Pourdeyhimi, B.}} \yr{2019}  \at{Modeling polymer crystallization
  kinetics in the meltblowing process}.  \jt{Industrial \& Engineering
  Chemistry Research}  \bvol{59}~(1),  \pg{399--412}.

\bibitem[Hao {\em et~al.\/}(2016)Hao, Zhou, Zhou, Che, Chu \&
  Wang]{hao2016dynamic}
{\sc \au{Hao, C.}, \au{Zhou, Y.}, \au{Zhou, X.}, \au{Che, L.}, \au{Chu, B.} \&
  \au{Wang, Z.}} \yr{2016}  \at{Dynamic control of droplet jumping by tailoring
  nanoparticle concentrations}.  \jt{Applied Physics Letters}  \bvol{109}~(2),
  \pg{021601}.

\bibitem[Harlow \& Shannon(1967)]{harlow1967splash}
{\sc \au{Harlow, F.~H.} \& \au{Shannon, J.~P.}} \yr{1967}  \at{The splash of a
  liquid drop}.  \jt{J. Appl. Phys.}  \bvol{38}~(10),  \pg{3855--3866}.

\bibitem[Josserand \& Thoroddsen(2016)]{josserand2016drop}
{\sc \au{Josserand, C.} \& \au{Thoroddsen, S.~T.}} \yr{2016}  \at{Drop impact
  on a solid surface}.  \jt{Annu. Rev. Fluid Mech.}  \bvol{48},  \pg{365--391}.

\bibitem[Kadoura \& Chandra(2013)]{kadoura2013rupture}
{\sc \au{Kadoura, M.} \& \au{Chandra, S.}} \yr{2013}  \at{Rupture of thin
  liquid films sprayed on solid surfaces}.  \jt{Exp. Fluids}  \bvol{54}~(2),
  \pg{1--11}.

\bibitem[Kittel {\em et~al.\/}(2018)Kittel, Roisman \&
  Tropea]{Kittelsplash2018}
{\sc \au{Kittel, H.~M.}, \au{Roisman, I.~V.} \& \au{Tropea, C.}} \yr{2018}
  \at{Splash of a drop impacting onto a solid substrate wetted by a thin film
  of another liquid}.  \jt{Phys. Rev. Fluids}  \bvol{3}~(073601).

\bibitem[Kolmogorov(1962)]{kolmogorov1962refinement}
{\sc \au{Kolmogorov, A.~N.}} \yr{1962}  \at{A refinement of previous hypotheses
  concerning the local structure of turbulence in a viscous incompressible
  fluid at high reynolds number}.  \jt{Journal of Fluid Mechanics}
  \bvol{13}~(1),  \pg{82--85}.

\bibitem[Landau \& Lifshitz(2013)]{landau2013statistical}
{\sc \au{Landau, L.~D} \& \au{Lifshitz, E.~M.}} \yr{2013} {\em Statistical
  Physics: Volume 5\/}.  \publ{Oxford: Butterworth-Heinemann}.

\bibitem[Layani {\em et~al.\/}(2014)Layani, Berman \&
  Magdassi]{layani2014printing}
{\sc \au{Layani, M.}, \au{Berman, R.} \& \au{Magdassi, S.}} \yr{2014}
  \at{Printing holes by a dewetting solution enables formation of a transparent
  conductive film}.  \jt{ACS Appl. Mater. Interfaces}  \bvol{6}~(21),
  \pg{18668--18672}.

\bibitem[Levin \& Hobbs(1971)]{levin1971splashing}
{\sc \au{Levin, Z.} \& \au{Hobbs, P.~V.}} \yr{1971}  \at{Splashing of water
  drops on solid and wetted surfaces: hydrodynamics and charge separation}.
  \jt{Phil. Trans. R. Soc. Lond. A}  \bvol{269}~(1200),  \pg{555--585}.

\bibitem[Lhuissier {\em et~al.\/}(2013)Lhuissier, Sun, Prosperetti \&
  Lohse]{lhuissier2013drop}
{\sc \au{Lhuissier, H.}, \au{Sun, C.}, \au{Prosperetti, A.} \& \au{Lohse, D.}}
  \yr{2013}  \at{Drop fragmentation at impact onto a bath of an immiscible
  liquid}.  \jt{Phys. Rev. Lett.}  \bvol{110}~(26),  \pg{264503}.

\bibitem[Macklin \& Metaxas(1976)]{macklin1976splashing}
{\sc \au{Macklin, W.~C.} \& \au{Metaxas, G.~J.}} \yr{1976}  \at{Splashing of
  drops on liquid layers}.  \jt{J. Appl. Phys.}  \bvol{47}~(9),
  \pg{3963--3970}.

\bibitem[Marengo {\em et~al.\/}(2011)Marengo, Antonini, Roisman \&
  Tropea]{marengo2011drop}
{\sc \au{Marengo, M.}, \au{Antonini, C.}, \au{Roisman, I.~V.} \& \au{Tropea,
  C.}} \yr{2011}  \at{Drop collisions with simple and complex surfaces}.
  \jt{Current Opinion in Colloid \& Interface Science}  \bvol{16}~(4),
  \pg{292--302}.

\bibitem[Pope(2001)]{pope2001turbulent}
{\sc \au{Pope, S.~B.}} \yr{2001} {\em Turbulent Flows\/}.  \publ{Cambridge:
  Cambridge University Press}.

\bibitem[Prunet-Foch {\em et~al.\/}(1998)Prunet-Foch, Legay, Vignes-Adler \&
  Delmotte]{prunet1998impacting}
{\sc \au{Prunet-Foch, B.}, \au{Legay, F.}, \au{Vignes-Adler, M.} \&
  \au{Delmotte, C.}} \yr{1998}  \at{Impacting emulsion drop on a steel plate:
  influence of the solid substrate}.  \jt{J. Colloid Interface Sci.}
  \bvol{199}~(2),  \pg{151--168}.

\bibitem[Rioboo {\em et~al.\/}(2003)Rioboo, Bauthier, Conti, Voue \&
  De~Coninck]{rioboo2003experimental}
{\sc \au{Rioboo, R.}, \au{Bauthier, C.}, \au{Conti, J.}, \au{Voue, M.} \&
  \au{De~Coninck, J.}} \yr{2003}  \at{Experimental investigation of splash and
  crown formation during single drop impact on wetted surfaces}.
  \jt{Experiments in Fluids}  \bvol{35}~(6),  \pg{648--652}.

\bibitem[Rioboo {\em et~al.\/}(2001)Rioboo, Tropea \&
  Marengo]{rioboo2001outcomes}
{\sc \au{Rioboo, R.}, \au{Tropea, C.} \& \au{Marengo, M.}} \yr{2001}
  \at{Outcomes from a drop impact on solid surfaces}.  \jt{Atom. Sprays}
  \bvol{11}~(2),  \pg{155--165}.

\bibitem[Roisman(2009)]{roisman2009inertia}
{\sc \au{Roisman, I.~V.}} \yr{2009}  \at{Inertia dominated drop collisions. ii.
  an analytical solution of the {N}avier--{S}tokes equations for a spreading
  viscous film}.  \jt{Phys. Fluids}  \bvol{21}~(5),  \pg{052104}.

\bibitem[Roisman {\em et~al.\/}(2007)Roisman, Gambaryan-Roisman, Kyriopoulos,
  Stephan \& Tropea]{roisman2007breakup}
{\sc \au{Roisman, I.~V.}, \au{Gambaryan-Roisman, T.}, \au{Kyriopoulos, O.},
  \au{Stephan, P.} \& \au{Tropea, C.}} \yr{2007}  \at{Breakup and atomization
  of a stretching crown}.  \jt{Physical Review E}  \bvol{76}~(2),  \pg{026302}.

\bibitem[Roisman {\em et~al.\/}(2008)Roisman, van Hinsberg \&
  Tropea]{roisman2008propagation}
{\sc \au{Roisman, I.~V.}, \au{van Hinsberg, N.~P.} \& \au{Tropea, C.}}
  \yr{2008}  \at{Propagation of a kinematic instability in a liquid layer:
  Capillary and gravity effects}.  \jt{Phys. Rev. E}  \bvol{77}~(4),
  \pg{046305}.

\bibitem[Sreenivasan(1995)]{sreenivasan1995universality}
{\sc \au{Sreenivasan, K~R}} \yr{1995}  \at{On the universality of the
  {K}olmogorov constant}.  \jt{Physics of Fluids}  \bvol{7}~(11),
  \pg{2778--2784}.

\bibitem[Stauffer({1979})]{Stauffer1979}
{\sc \au{Stauffer, D.}} \yr{{1979}}  \at{Scaling theory of percolation
  clusters}.  \jt{Physics Reports}  \bvol{{54}}~({1}),  \pg{{1 -- 74}}.

\bibitem[Stauffer(1985)]{Stauffer1985}
{\sc \au{Stauffer, D.}} \yr{1985} {\em Introduction to Percolation Theory\/}.
  \publ{Taylor and Francis}.

\bibitem[Taylor(1959)]{Taylor1959}
{\sc \au{Taylor, G.~I.}} \yr{1959}  \at{The dynamics of thin sheets of fluid.
  iii. disintegration of fluid sheets.}  \jt{Proc. R. Soc. Lond. A}
  \bvol{253},  \pg{625--639}.

\bibitem[Taylor \& Michael(1973)]{Taylor1973}
{\sc \au{Taylor, G.~I.} \& \au{Michael, D.~H.}} \yr{1973}  \at{On making holes
  in a sheet of liquid}.  \jt{J. Fluid Mech.}  \bvol{58}~(4),  \pg{625--639}.

\bibitem[Wakimoto \& Azuma(2009)]{wakimoto2009influence}
{\sc \au{Wakimoto, T.} \& \au{Azuma, T.}} \yr{2009}  \at{Influence of the
  physical properties of a liquid on perforations in a radial liquid sheet
  jet}.  \jt{Journal of Fluid Science and Technology}  \bvol{4}~(2),
  \pg{359--367}.

\bibitem[Wang \& Chen(2000)]{wang2000splashing}
{\sc \au{Wang, A.-B.} \& \au{Chen, C.-C.}} \yr{2000}  \at{Splashing impact of a
  single drop onto very thin liquid films}.  \jt{Phys. Fluids}  \bvol{12}~(9),
  \pg{2155--2158}.

\bibitem[Worthington \& Cole(1897)]{worthington1897impact}
{\sc \au{Worthington, A.~M.} \& \au{Cole, R.~S.}} \yr{1897}  \at{Impact with a
  liquid surface, studied by the aid of instantaneous photography}.  \jt{Phil.
  Trans. R. Soc. Lond. A}  \bvol{189},  \pg{137--148}.

\bibitem[Yang {\em et~al.\/}(2005)Yang, Liu \&
  Shivpuri]{yang2005physiothermodynamics}
{\sc \au{Yang, L.}, \au{Liu, C.} \& \au{Shivpuri, R.}} \yr{2005}
  \at{Physiothermodynamics of lubricant deposition on hot die surfaces}.
  \jt{CIRP Ann-Manuf. Technol.}  \bvol{54}~(1),  \pg{253--256}.

\bibitem[Yarin(1992)]{yarin1992flow}
{\sc \au{Yarin, A.~L.}} \yr{1992}  \at{Flow-induced on-line crystallization of
  rodlike molecules in fiber spinning}.  \jt{Journal of Applied Polymer
  Science}  \bvol{46}~(5),  \pg{873--878}.

\bibitem[Yarin(1993)]{yarin1993free}
{\sc \au{Yarin, A.~L.}} \yr{1993} {\em Free {L}iquid {J}ets and {F}ilms:
  {H}ydrodynamics and {R}heology\/}.  \publ{Harlow Longman and J. Wiley \&
  Sons}.

\bibitem[Yarin(2006)]{yarin2006drop}
{\sc \au{Yarin, A.~L.}} \yr{2006}  \at{Drop impact dynamics: splashing,
  spreading, receding, bouncing ...}  \jt{Annu. Rev. Fluid Mech.}  \bvol{38},
  \pg{159--192}.

\bibitem[Yarin {\em et~al.\/}(2017)Yarin, Roisman \&
  Tropea]{CollisionPhenomena2017}
{\sc \au{Yarin, A.~L.}, \au{Roisman, I.~V.} \& \au{Tropea, C.}} \yr{2017} {\em
  Collision Phenomena in Liquids and Solids\/}.  \publ{Cambridge: Cambridge
  University Press}.

\bibitem[Yarin \& Weiss(1995)]{yarin1995impact}
{\sc \au{Yarin, A.~L.} \& \au{Weiss, D.~A.}} \yr{1995}  \at{Impact of drops on
  solid surfaces: self-similar capillary waves, and splashing as a new type of
  kinematic discontinuity}.  \jt{J. Fluid Mech.}  \bvol{283},  \pg{141--173}.

\end{thebibliography}

\end{document}